\documentclass[12pt,dvips]{article}
\usepackage{rotating}
\usepackage{epsfig}
\usepackage{color}
\usepackage{cite}
\usepackage{amsmath}

\newcommand{\lsim}{\mathrel{\mathop{\kern 0pt \rlap
  {\raise.2ex\hbox{$<$}}}
  \lower.9ex\hbox{\kern-.190em $\sim$}}}
\newcommand{\gsim}{\mathrel{\mathop{\kern 0pt \rlap
  {\raise.2ex\hbox{$>$}}}
  \lower.9ex\hbox{\kern-.190em $\sim$}}}

\newcommand{\Vac}[1]{\bigg\langle{#1}\bigg\rangle}

\def\theequation{\arabic{section}.\arabic{equation}}

\begin{document}

\def\thefootnote{\fnsymbol{footnote}}

\begin{flushright}
{\tt arXiv:1006.1458 [hep-ph]}\\
June 2010
%Last modified by JSL on \today
%Last modified by E.S. on \today
%Last modified by kc on \today
\end{flushright}

\begin{center}
{\bf {\LARGE
The Higgs Boson Sector of  the\\[3mm]
Next-to-MSSM with CP Violation 
} }
%\\[3.mm]
\end{center}

\medskip

\begin{center}{\large
Kingman Cheung$^{a,b,c}$,
Tie-Jiun Hou$^{a,b}$,
Jae~Sik~Lee$^b$, and
Eibun Senaha$^b$ }
\end{center}

\begin{center}
{\em $^a$ Department of Physics, National Tsing Hua University, Hsinchu, Taiwan 300}\\[0.2cm]
{\em $^b$Physics Division, National Center for Theoretical Sciences,
Hsinchu, Taiwan 300}\\[0.2cm]
{\em $^c$Division of Quantum Phases \& Devices,
                  Konkuk University,  Seoul 143-701, Korea}\\[0.2cm]
%{\em $^d$Department of Physics and Center for Mathematics and
%  Theoretical Physics,}\\
%{\em National Central University, Chung-Li, Taiwan 32054}
\end{center}

\bigskip\bigskip

\begin{abstract}
We perform a comprehensive study of the Higgs sector in the framework
of the next-to-minimal supersymmetric standard model with 
CP-violating parameters in the superpotential and in the
soft-supersymmetry-breaking sector.  Since the CP is no longer a good 
symmetry, the two CP-odd and the three CP-even Higgs bosons of 
the next-to-minimal supersymmetric standard model in the CP-conserving limit
will mix.  We show explicitly how the mass spectrum and couplings
to gauge bosons of the various Higgs bosons change when the CP-violating
phases take on nonzero values.  
We include full one-loop 
and the logarithmically enhanced two-loop effects employing the
renormalization-group (RG) improved approach.
In addition, the LEP limits, the global minimum condition, and the positivity of
the square of the Higgs-boson mass have been imposed. We demonstrate the
effects on the Higgs-mass spectrum and the couplings to gauge bosons
with and without the RG-improved corrections.  
Substantial modifications
to the allowed parameter space happen because of the changes to the
Higgs-boson spectrum and their couplings
with the RG-improved corrections. Finally, we 
calculate the mass spectrum and couplings of the few selected
scenarios and compare to the previous results in literature where possible;
in particular, we illustrate a scenario motivated by electroweak
baryogenesis.  
\end{abstract}
\newpage 
%
%%%%%%%%%%%%%%%%%%%%%-------------------
\section{Introduction}
%%%%%%%%%%%%%%%%%%%%%-------------------
Supersymmetry (SUSY) is the leading candidate for the physics beyond the
standard model (SM).  It not only solves the gauge hierarchy problem,
but also provides a dynamical mechanism for electroweak symmetry
breaking and a natural candidate for the dark matter. 
The minimal supersymmetric extension of the SM (MSSM)
has attracted much phenomenological and theoretical interest but 
it suffers from the so-called 
little hierarchy problem and the $\mu$ problem.

An extension with an extra singlet superfield, known as the
next-to-minimal supersymmetric standard model 
(NMSSM)~\cite{NMSSM:0,NMSSM:1,NMSSM:ECPV,Degrassi:2009yq,NMSSM:review}
was motivated to
provide a natural solution to the $\mu$ problem.  The $\mu$ parameter
in the term $\mu H_u H_d$ of the superpotential of the MSSM naturally
has its value at either $M_{\rm Planck}$ or zero (due to a symmetry).
However, the radiative electroweak symmetry breaking conditions
require the $\mu$ parameter to be of the same order as the $Z$-boson mass
for fine-tuning reasons. Such a conflict was coined as 
the $\mu$ problem~\cite{mu-problem}.
In the NMSSM, the $\mu$ term is generated dynamically
through the vacuum-expectation value (VEV), $v_S$, of the scalar
component of the additional Higgs field $S$, which is naturally of the
order of the SUSY-breaking scale.  Thus, an effective $\mu$ parameter
of the order of the electroweak scale is generated.
The NMSSM was recently revived because it was shown that it can
effectively relieve the little hierarchy problem \cite{derm}.  Because of
the additional Higgs singlet field and an approximate Peccei-Quinn (PQ) symmetry, the
NMSSM naturally has a light pseudoscalar Higgs boson $a_1$.  It has been
shown \cite{derm} that, in most parameter space that is natural, the
SM-like Higgs boson can decay into a pair of light pseudoscalar bosons
with a branching ratio larger than $0.7$.  Thus, the branching ratio
of the SM-like Higgs boson into $b\bar b$ would be less than $0.3$ and
so the LEPII bound is effectively reduced to around 100 GeV
\cite{LEP2003}.  Since the major decay modes of the Higgs boson are
no longer $b\bar b$, unusual search modes have been investigated
\cite{nmssm-new}.

In SUSY models, CP-violating phases naturally appear in the $\mu$ term of 
the superpotential and in the soft-SUSY breaking terms.
The nonobservation of electric dipole moments (EDMs) for
thallium~\cite{Regan:2002ta}, neutron~\cite{Baker:2006ts},
and mercury~\cite{Romalis:2000mg,Griffith:2009zz} is known to
constrain CP-violating phases very tightly. Nevertheless,
cancellations among various contributions may occur 
among several contributions to the three measured EDMs,
thus still allowing sizable CP phases~\cite{Ibrahim:1998je,Ellis:2008zy}.
In the MSSM,
the nonvanishing CP phases could radiatively induce 
significant mixing between the CP-even and 
CP-odd states~\cite{CPmixing0,CPmixing1,CPmixing1.5,CPmixing2},
giving rise to a number of interesting CP-violating phenomena
and substantial modifications to 
Higgs-boson phenomenology~\cite{Lee:2008eqa,Accomando:2006ga}.
In particular, the lightest Higgs boson can be as light as a few GeV
with almost vanishing couplings to the weak gauge bosons
when the CP-violating phases are maximal.  The decay patterns of the
heavier Higgs bosons become much more 
complicated compared to the CP-conserving case
because of the loss of its CP parities~\cite{CPH_decay,cpsuperh}.
These combined features make the Higgs boson-searches at LEP difficult;
consequently, the Higgs boson lighter than $\sim$ 50 GeV
can survive the LEP limit~\cite{Schael:2006cr}.

In this work, we study the NMSSM Higgs sector with CP violation.
In the CP-conserving limit, the neutral Higgs sector in 
the NMSSM includes two CP-odd and three CP-even states.
With CP violation the 5 neutral Higgs bosons lose their CP parities
and all mix together.
We anticipate a whole new set of phenomena
associated with the singlet extension of the MSSM in the presence of 
nontrivial CP-violating phases in the VEVs, and the  $\mu$ and 
soft SUSY-breaking parameters.
As the first step toward this new extension, we calculate 
the whole mass spectrum
of the Higgs sector as well as the couplings to the vector gauge bosons,
which will dictate the production and decay patterns of the
Higgs bosons.  Phenomenology associated with the 
CP-violating NMSSM Higgs sector will be performed
in future works. We list a few possible directions at the end of
the Conclusions.

We include in this calculation the important corrections to the Higgs spectrum
in order to have more precise comparisons to current experimental limits:
(i) full one-loop corrections to the Higgs-boson masses,
and 
(ii) logarithmically enhanced two-loop corrections of order 
${\cal O}(g_s^2 h^4)$ 
and ${\cal O}(h^6)$ with the renormalization-group (RG) improvement 
and minimization of the two-loop corrections.
We also impose highly desirable conditions to limit the parameter space:
(i) the LEP limits on the Higgs-boson mass and 
the chargino mass limit,
(ii) the global minimum condition---the local minimum that we obtain is 
indeed the global minimum, and
(iii) the positivity of the square of the Higgs-boson masses.
We found that the RG-improved corrections have significant reduction
in the allowed parameter space with respect to the LEP limits, the 
global minimum condition, and the positivity of the Higgs-mass squared.

Before we close this section let us list the parameters of this study.
We have (i) the usual soft parameters in the MSSM: sfermion masses,
$A$ parameters, soft Higgs-boson masses, and $\tan\beta$;
(ii) the additional parameters arisen in NMSSM: $\lambda$ and $\kappa$
in the superpotential, $A_\lambda$ and $A_\kappa$ in the soft-breaking sector,
and the VEV $v_S$ of the singlet Higgs field; (iii) the CP phases of
$A$ parameters, $\lambda$, $\kappa$, and the VEVs (but not all independent).

The organization of this paper is as follows. We write down the formalism in details in 
the next section, including minimization conditions, combinations of
CP phases, tree-level mass matrices, and mixing between scalar and
pseudoscalar Higgs bosons. In Sec. III, we calculate the full one-loop
and logarithmically enhanced two-loop corrections using the
RG-improved approach.  Numerical presentation for a number of 
interesting scenarios will be demonstrated in Sec. IV, including
comparisons to the previous results in literature where possible. 
We conclude and discuss our results in Sec. V.

%%%%%%%%%%%%%%%%%%%%%-------------------
\setcounter{equation}{0}
\section{Higgs sector at the tree level}
%%%%%%%%%%%%%%%%%%%%%-------------------

%-------------------
\subsection{Conventions}
%-------------------

To begin with, we first introduce the NMSSM superpotential:
\begin{equation}
  \label{Wpot}
W_{\rm NMSSM}\ =\ \widehat{U}^C {\bf h}_u \widehat{Q} \widehat{H}_u\:
+\:   \widehat{D}^C {\bf h}_d \widehat{H}_d \widehat{Q}  \: +\:
\widehat{E}^C {\bf h}_e \widehat{H}_d \widehat{L} \: +\: 
\lambda \widehat{S} \widehat{H}_u \widehat{H}_d\ \: + \:
\frac{\kappa}{3}\ \widehat{S}^3 \ ,
\end{equation}
where $\widehat{S}$ denotes the singlet Higgs superfield,
$\widehat{H}_{u,d}$  are the two SU(2)$_L$ doublet Higgs superfields, and
$\widehat{Q}$,  $\widehat{L}$ and $\widehat{U}^C$,  $\widehat{D}^C$,
$\widehat{E}^C$ are the matter doublet and singlet superfields, respectively, 
related to up- and  down-type quarks and  charged leptons.  
We note that, especially, the last cubic term 
with a dimensionless coupling $\kappa$
respects an extra discrete $Z_3$ symmetry.
The  Yukawa couplings
${\bf  h}_{u,d,e}$ are  $3\times  3$ complex  matrices describing  the
quark and charged-lepton masses and mixing. In the expression, for example,
the notation 
$\widehat{H}_u \widehat{H}_d \equiv \epsilon_{\alpha\beta}
(\widehat{H}_u)^\alpha (\widehat{H}_d)^\beta$ is implicit.
The superpotential leads to the tree-level Higgs potential, which
is given by the sum
\begin{eqnarray}
V_0=V_F+V_D+V_{\rm soft},
\end{eqnarray}
where each term is given by
\begin{eqnarray}
V_F&=&|\lambda|^2|S|^2(H_d^\dagger H_d+H_u^\dagger H_u)
        +|\lambda H_u H_d+\kappa S^2|^2,\nonumber \\
V_D&=&\frac{g_2^2+g_1^2}{8}(H_d^\dagger H_d-H_u^\dagger H_u)^2
        +\frac{g_2^2}{2}(H_d^\dagger H_u)(H_u^\dagger H_d),\nonumber \\
V_{\rm soft}&=&m_1^2 H_d^\dagger H_d+m_2^2 H_u^\dagger H_u
        +m_S^2|S|^2
        +\left(\lambda A_{\lambda}S H_u H_d
        -\frac{1}{3}\kappa A_\kappa S^3+{\rm h.c.}\right),
\end{eqnarray}
with the gauge-coupling constants $g_1=e/\cos\theta_W$ and $g_2=e/\sin\theta_W$.
Note the unusual minus($-$) sign for the singlet soft-trilinear term proportional
to $A_\kappa$\footnote{One can go back to the usual convention
by taking $R_\kappa \to -R_\kappa$
or $\phi^\prime_\kappa \to \phi^\prime_\kappa+\pi$ in below.}.

Parametrizing the component fields of
the two doublet and one singlet scalar Higgs fields and 
the VEVs as follows,
\begin{eqnarray}
H_d&=&
\hphantom{e^{i\theta}}\,
\left(
\begin{array}{c}
\frac{1}{\sqrt{2}}\,(v_d+\phi^0_d+ia_d) \\
\phi_d^-
\end{array}
%\right), \quad
\right), \nonumber \\[1mm]
%H_u=
H_u&=&
e^{i\theta}\,\left(
\begin{array}{c}
\phi_u^+\\
\frac{1}{\sqrt{2}}\,(v_u+\phi^0_u+ia_u)
\end{array}
\right), \nonumber \\[1mm]
S&=&\frac{e^{i\varphi}}{\sqrt{2}}\,(v_S+\phi^0_S+ia_S)\,,
\label{eq:higgsparam}
\end{eqnarray}
we get the following tadpole conditions for the presumed vacuum:
\begin{eqnarray}
\frac{1}{v_d}\Vac{\frac{\partial V_0}{\partial \phi^0_d}}
&=&m_{1}^{2}+\frac{g_{2}^{2}+g_{1}^{2}}{8}(v_{d}^{2}-v_{u}^{2})-R_\lambda\frac{v_uv_S}{v_d}
        +\frac{|\lambda|^2}{2}(v_u^2+v_S^2)-\frac{1}{2}\mathcal{R}\frac{v_uv^2_S}{v_d}=0,
\nonumber \\
\frac{1}{v_u}\Vac{\frac{\partial V_0}{\partial \phi^0_u}}
&=&m_{2}^{2}-\frac{g_{2}^{2}+g_{1}^{2}}{8}(v_{d}^{2}-v_{u}^{2})-R_\lambda\frac{v_dv_S}{v_u}
        +\frac{|\lambda|^2}{2}(v_d^2+v_S^2)-\frac{1}{2}\mathcal{R}\frac{v_dv^2_S}{v_u}=0,
\nonumber \\
\frac{1}{v_S}\Vac{\frac{\partial V_0}{\partial \phi^0_S}}
&=&m_S^2-R_\lambda\frac{v_dv_u}{v_S}+\frac{|\lambda|^2}{2}(v_d^2+v_u^2)
        +|\kappa|^2v^2_S-\mathcal{R}v_dv_u-R_\kappa v_S,
\label{eq:cpeventad} \\
\frac{1}{v_u}\Vac{\frac{\partial V_0}{\partial a_d}}
&=&\frac{1}{v_d}\Vac{\frac{\partial V_0}{\partial a_u}}
        =I_\lambda v_S+\frac{1}{2}\mathcal{I}v^2_S=0,
\nonumber \\
\frac{1}{v_S}\Vac{\frac{\partial V_0}{\partial a_S}}
&=&I_\lambda \frac{v_dv_u}{v_S}-\mathcal{I}v_dv_u+I_\kappa v_S=0\,,
\label{eq:cpoddtad}
\end{eqnarray}
where
\begin{eqnarray}
\mathcal{R} &=& |\lambda| |\kappa|\, \cos(\phi^\prime_\lambda-\phi^\prime_\kappa)\,,
\hspace{1.0cm}
\mathcal{I} = |\lambda| |\kappa|\, \sin(\phi^\prime_\lambda-\phi^\prime_\kappa)\,,
\nonumber \\
R_\lambda &=& \frac{|\lambda| |A_\lambda|}{\sqrt{2}}\,
\cos(\phi^\prime_\lambda+\phi_{A_\lambda})\,,
\ \ \
R_\kappa = \frac{|\kappa| |A_\kappa|}{\sqrt{2}}\,
\cos(\phi^\prime_\kappa+\phi_{A_\kappa})\,,
\end{eqnarray}
with
\begin{equation}
\phi^\prime_\lambda \equiv \phi_\lambda+\theta+\varphi \ \ \ {\rm and} \ \ \
\phi^\prime_\kappa \equiv \phi_\kappa+3\varphi\,.
\end{equation}
The other parameters $I_\lambda$ and $I_\kappa$ can be reexpressed in terms of $\mathcal{I}$
using the CP-odd tadpole conditions as
\begin{eqnarray}
I_\lambda &=& \frac{|\lambda| |A_\lambda|}{\sqrt{2}}\,
\sin(\phi^\prime_\lambda+\phi_{A_\lambda}) = - \frac{1}{2}\,\mathcal{I}\, v_S\,,
\nonumber \\
I_\kappa &=& \frac{|\kappa| |A_\kappa|}{\sqrt{2}}\,
\sin(\phi^\prime_\kappa+\phi_{A_\kappa}) = \frac{3}{2}\,\mathcal{I}\, \frac{v_dv_u}{v_S}\,.
\label{eq:ilik}
\end{eqnarray}
Therefore, the only rephasing invariant physical CP phase at the tree level
is $\phi^\prime_\lambda-\phi^\prime_\kappa$ and, once the absolute values
of $|\lambda|$, $|\kappa|$, $|A_\lambda|$, and $|A_\kappa|$ are given, the other 
two combinations of CP phases, $\phi^\prime_\lambda+\phi_{A_\lambda}$
and $\phi^\prime_\kappa+\phi_{A_\kappa}$, can be determined  up to a 
twofold ambiguity 
using the two CP-odd tadpole conditions in Eq.~(\ref{eq:cpoddtad}).
We also observe that the soft masses $m_1^2$, $m_2^2$, and $m_S^2$ can be removed using the
three CP-even tadpole conditions in Eq.~(\ref{eq:cpeventad}).

From the potential, with the parametrization of the scalar Higgs fields as
in Eq.~(\ref{eq:higgsparam}), the scalar mass terms can be derived and
they can be cast into the form
\begin{equation}
-{\cal L}_{\rm mass} = M_{H^\pm}^{(0)\,2} H^+ H^- +
\frac{1}{2}\, \Phi^T\, {\cal M}_N^{(0)\,2}\, \Phi
\end{equation}
where $H^\pm=\phi_d^\pm \sin\beta + \phi_u^\pm \cos\beta$ with $\tan\beta = v_u/v_d$ and
\begin{equation}
\Phi^T\equiv\left(
\phi^0_d\,, \phi^0_u\,, \phi^0_S\,, a\,, a_S \right)
\end{equation}
with $a=a_d \sin\beta + a_u \cos\beta$, rotating away the zero mass Goldstone states.
The tree-level charged Higgs-boson mass is given by
\begin{eqnarray}
M_{H^\pm}^{(0)\,2}
        =M_W^2+(2R_\lambda+\mathcal{R}v_S)\frac{v_S}{\sin2\beta}
        -\frac{|\lambda|^2}{2}v^2.\label{eq:mch1}
\end{eqnarray}
The $5\times 5$ symmetric mass matrix for the neutral Higgs boson is given by
\begin{eqnarray}
\hspace{-1.0cm}
{\cal M}^{(0)\,2}_N=
\left(
        \begin{array}{ccccc}
\left(\mathcal{M}^{(0)\,2}_S\right)_{11} &
\left(\mathcal{M}^{(0)\,2}_S\right)_{12} &
\left(\mathcal{M}^{(0)\,2}_S\right)_{13}
        & 0 & -\frac{3}{2}\mathcal{I}v_uv_S \\
\left(\mathcal{M}^{(0)\,2}_S\right)_{21} &
\left(\mathcal{M}^{(0)\,2}_S\right)_{22} &
\left(\mathcal{M}^{(0)\,2}_S\right)_{23}
        & 0 & -\frac{3}{2}\mathcal{I}v_dv_S \\
\left(\mathcal{M}^{(0)\,2}_S\right)_{31} &
\left(\mathcal{M}^{(0)\,2}_S\right)_{32} &
\left(\mathcal{M}^{(0)\,2}_S\right)_{33}
        & \frac{1}{2}\mathcal{I}vv_S &  2\mathcal{I}v_dv_u \\
        0 & 0 & \frac{1}{2}\mathcal{I}vv_S
        & (R_\lambda+\frac{1}{2}\mathcal{R}v_S)\frac{v^2v_S}{v_dv_u}
        & (R_\lambda -\mathcal{R}v_S)v \\
        -\frac{3}{2}\mathcal{I}v_uv_S & -\frac{3}{2}\mathcal{I}v_dv_S
        & 2\mathcal{I}v_dv_u
        & (R_\lambda -\mathcal{R}v_S)v
        & R_\lambda\frac{v_dv_u}{v_S}+2\mathcal{R}v_dv_u+3R_\kappa v_S
        \end{array}
\right)\,,
\label{eq:mNH}
\end{eqnarray}
with
\begin{eqnarray}
\left({\cal M}^{(0)\,2}_S\right)_{11}&=&\frac{g_2^2+g_1^2}{4}v_d^2
        +\left(R_\lambda+\frac{1}{2}\mathcal{R}v_S\right)\frac{v_uv_S}{v_d}, \nonumber \\
\left({\cal M}^{(0)\,2}_S\right)_{22}&=&\frac{g_2^2+g_1^2}{4}v_u^2
        +\left(R_\lambda+\frac{1}{2}\mathcal{R}v_S\right)\frac{v_dv_S}{v_u}, \nonumber \\
\left({\cal M}^{(0)\,2}_S\right)_{33}&=&R_\lambda\frac{v_dv_u}{v_S}+2|\kappa|^2v^2_S-R_\kappa
v_S, \nonumber \\
\left({\cal M}^{(0)\,2}_S\right)_{12}&=&\left({\cal M}^{(0)\,2}_S\right)_{21}
        =\left(-\frac{g_2^2+g_1^2}{4}+|\lambda|^2\right)v_dv_u
        -\bigg(R_\lambda+\frac{1}{2}\mathcal{R}v_S\bigg)v_S, \nonumber \\
\left({\cal M}^{(0)\,2}_S\right)_{13}&=&\left({\cal M}^{(0)\,2}_S\right)_{31}
        =-R_\lambda v_u+|\lambda|^2v_dv_S-\mathcal{R}v_uv_S,\nonumber \\
\left({\cal M}^{(0)\,2}_S\right)_{23}&=&\left({\cal M}^{(0)\,2}_S\right)_{32}
        =-R_\lambda v_d+|\lambda|^2v_uv_S-\mathcal{R}v_dv_S.
\end{eqnarray}
%

%-------------------
\subsection{Mixing and mass spectrum with ${\cal I} \neq 0$}
%-------------------
When ${\cal I} \neq 0$, we should consider the full $5\times 5$ matrix (\ref{eq:mNH})
for the neutral Higgs-boson masses and mixing. In this case,
the neutral Higgs bosons do not have to carry any definite CP parities
and its mixing is described by 
the orthogonal $5\times 5$ matrix $O_{\alpha i}$ as
\begin{equation}
%\Phi = 
\left( \phi^0_d\,, \phi^0_u\,, \phi^0_S\,, a\,, a_S \right)^T \ = \
O_{\alpha i} 
\left( H_1\,, H_2\,, H_3\,, H_4\,, H_5 \right)^T
\end{equation}
with $H_{1(5)}$  the lightest (heaviest) Higgs-mass eigenstate. Because of
its CP-violating mixing, the couplings of the neutral Higgs bosons to the
SM and SUSY particles are significantly modified. Among them
the most eminent one is the couplings of the Higgs bosons to 
weak gauge bosons in the interaction Lagrangian:
\begin{eqnarray}
{\cal L}_{HVV} & = & g_2\,M_W \, \left(W^+_\mu W^{- \mu}\ + \
\frac{1}{2c_W^2}\,Z_\mu Z^\mu\right) \, \sum_i \,g_{_{H_iVV}}\, H_i
\,,\\[3mm]
{\cal L}_{HHZ} &=& \frac{g_2}{4c_W} \sum_{i,j} g_{_{H_iH_jZ}}\, Z^{\mu}
(H_i\, \!\stackrel {\leftrightarrow} {\partial}_\mu H_j) \,, \\ [3mm]
{\cal L}_{HH^\pm W^\mp} &=& -\frac{g_2}{2} \, \sum_i \, g_{_{H_iH^+
W^-}}\, W^{-\mu} (H_i\, i\!\stackrel{\leftrightarrow}{\partial}_\mu
H^+)\, +\, {\rm h.c.}\,,
\end{eqnarray}
where the couplings $g_{_{H_iVV}}$, $g_{_{H_iH_jZ}}$ and $g_{_{H_iH^+
W^-}}$ are given in terms of the neutral Higgs-boson mixing matrix $O$
by
\begin{eqnarray}
g_{_{H_iVV}} &=& c_\beta\, O_{1 i}\: +\: s_\beta\, O_{2 i}
\, ,\nonumber \\
g_{_{H_iH_jZ}} &=& 
\left[ (O_{4i}\, (c_\beta\, O_{2j} - s_\beta\, O_{1j}) - (i \leftrightarrow j)
\right]
\nonumber \\
g_{_{H_iH^+ W^-}} &=& c_\beta\, O_{2 i} - s_\beta\, O_{1 i}
- i O_{4i} \, ,
\end{eqnarray}
leading to the following sum rules:
\begin{eqnarray}
\sum_{i=1}^5\, g_{_{H_iVV}}^2\ &=&\ 1\,,
\nonumber \\
\sum_{i>j}^5\, g_{_{H_iH_jZ}}^2\ &=&\ 1\,,
\nonumber \\
g_{_{H_iVV}}^2+|g_{_{H_iH^+ W^-}}|^2\ &=&\ 1-O_{3i}^2-O_{5i}^2\,\quad {\rm for~ each}~ i\,.
\end{eqnarray}

\medskip 

When the heavier two states have mass larger than that of the other three states, 
they effectively decouple from the mixing
and the lighter three states tend to mix among themselves. 
To have a better understanding in this case we introduce the 
following basis:
\begin{equation}
\Phi^\prime\equiv
\left( \phi^0_H \,, a\,, \phi^0_h \,, \phi^0_S \,, a_S  \right)^T=
\ U \ \Phi  =
\ U \ \left( \phi^0_d \,, \phi^0_u\,, \phi^0_S \,, a \,, a_S  \right)^T
\end{equation}
with
\begin{equation}
U = \left(\begin{array}{ccccc}
-s_\beta & c_\beta & 0 & 0 & 0 \\
 0 & 0  & 0 & 1 & 0 \\
 c_\beta & s_\beta & 0 & 0 & 0 \\
 0 & 0  & 1 & 0 & 0 \\
 0 & 0  & 0 & 0 & 1
\end{array} \right)\,.
\end{equation}
In the $\Phi^\prime$ basis, the symmetric $5 \times 5$ mass matrix takes the form
\begin{equation}
\left({\cal M}^{(0)\,2}_N\right)^\prime=
U \ {\cal M}^{(0)\,2}_N \ U^T =
\left(\begin{array}{cc}
 {\cal M}_J^2 & {\cal M}_{JL}^2 \\
 \left({\cal M}_{JL}^2\right)^T & {\cal M}_L^2  
\end{array} \right)\,,
\end{equation}
where
\begin{equation}
{\cal M}_J^2 = \left(\begin{array}{cc}
M_A^2+\left(M_Z^2-\frac{|\lambda|^2 v^2}{2}\right)\,s_{2\beta}^2 &
0 \\[0.2cm]
0 &
M_A^2
\end{array} \right)\,,
\end{equation}
with
\begin{eqnarray}
M_{A}^2&\equiv &
\frac{v_S}{\sin 2\beta}\,(2\,R_\lambda+v_S\,{\cal R}) \,.
\label{eq:ma}
\end{eqnarray}
Then, the tree-level charged  Higgs-boson mass can be rewritten as
\begin{equation}
M_{H^\pm}^{(0)\,2}
        =M_W^2+M_A^2-\frac{|\lambda|^2}{2}v^2.\label{eq:mch2}
\end{equation}
On the other hand, the $2\times 3$ heavy-light mixing matrix is
\begin{equation}
{\cal M}_{JL}^2 = \left(\begin{array}{ccc}
\left(-\frac{M_Z^2}{2}+\frac{|\lambda|^2 v^2}{4}\right)\,s_{4\beta} &
-\frac{v}{4v_S}\,M_A^2\,s_{4\beta}-\frac{vv_S}{2}\,{\cal R}\,c_{2\beta} &
-\frac{3}{2} v v_S \, {\cal I}\,c_{2\beta} \\[0.3cm]
0 &
\frac{1}{2} v v_S\,{\cal I} &
\frac{v}{2v_S}\,M_A^2\,s_{2\beta}-\frac{3\,v v_S}{2}\,{\cal R} 
\end{array} \right)\,,
\label{eq:MJL}
\end{equation}
and the symmetric $3\times 3$ mass matrix for the lighter states
is given by
\begin{equation}
{\cal M}_{L}^2 = \left(\begin{array}{ccc}
 M_Z^2\, c_{2\beta}^2 +\frac{1}{2}\, |\lambda|^2\, v^2\, s_{2\beta}^2 &
v v_S\left[\left(|\lambda|^2-\frac{M_A^2}{2v_S^2}\, s_{2\beta}^2\right)
-\frac{\cal R}{2}\,s_{2\beta} \right] &
-\frac{3}{2} v v_S \, {\cal I}\,s_{2\beta} \\[0.5cm]
 &
v^2\left(\frac{M_A^2}{4v_S^2}\,s^2_{2\beta} -\frac{\cal R}{4} s_{2\beta}\right)
&
v^2\,{\cal I}\,s_{2\beta} \\[0.2cm]
%-
 &
+v_S^2\,\left(2|\kappa|^2-\frac{R_\kappa}{v_S} \right)
&
 \\[0.5cm]
%-
 &
 &
\frac{v^2}{4v_S^2}\,M_A^2\,s_{2\beta}^2
+\frac{3\,v^2}{4}\,{\cal R}\,s_{2\beta}
\\
&
&
+3\,R_\kappa\,v_S 
\end{array} \right)\,.
\label{eq:ML}
\end{equation}
We observe that, in the leading order,
${\cal M}_J^2 \sim M_A^2$, ${\cal M}_{JL}^2 \sim \epsilon\,M_A^2 $, and
${\cal M}_L^2 \sim \epsilon^2\,M_A^2$ with
\begin{equation}
\epsilon = \max\left(
s_{2\beta}\,,\frac{v}{M_A}\,,\frac{v_S}{M_A}\,,\frac{R_\kappa}{M_A}
\right)\,.
\end{equation}
In this case, the mass matrix 
$\left({\cal M}^{(0)\,2}_N\right)^\prime$ could be 
systematically block diagonalized 
order by order in $\epsilon$, as described in Appendix A.
In the block-diagonalized basis, the
symmetric heavier-state mass matrix becomes
\begin{eqnarray}
\widetilde{ {\cal M}_J^2 } 
& = & \left(\begin{array}{cc}
M_A^2\left(1+\frac{v^2}{4v_S^2}\,s_{2\beta}^2\right) 
+\frac{v^2}{2}\,{\cal R}\,s_{2\beta}
& v^2\,{\cal I}\,s_{2\beta}
\\[0.5cm]
&
M_A^2\left(1+\frac{v^2}{4v_S^2}\,s_{2\beta}^2\right) 
-\frac{3v^2}{2}\,{\cal R}\,s_{2\beta}
\end{array} \right) \ + \ M_A^2\cdot {\cal O}(\epsilon^4)\,, \nonumber
\end{eqnarray}
leading to two almost degenerate eigenmasses
\begin{equation}
M^2_{J_1,J_2} 
\approx M_A^2\left(1+\frac{v^2}{4v_S^2}\,s_{2\beta}^2\right)
-v^2 s_{2\beta} \left({\cal R}/2\pm\sqrt{{\cal R}^2+{\cal I}^2}\right)
\end{equation}
and the mass splitting is of order ${\cal O}(\epsilon^3)$.
On the other hand, by defining 
\begin{equation}
Y\equiv M_A^2 s_{2\beta}^2-2|\lambda|^2 v_S^2 \,,
\end{equation}
one may obtain the following symmetric mass matrix for the lighter states
\begin{eqnarray}
\widetilde{ {\cal M}_L^2 } 
& =& \left(\begin{array}{ccc}
%---------
M_Z^2 &
-\frac{v Y}{2v_S}
-\frac{vv_S}{2}\,{\cal R}\,s_{2\beta} &
-\frac{3vv_S}{2}\,{\cal I}\,s_{2\beta} \\[0.5cm]
%---------
&
v_S^2\,\left(2|\kappa|^2-\frac{R_\kappa}{v_S}\right) 
-\frac{3v^2}{4}\,{\cal R}\,s_{2\beta}&
0 \\[0.5cm]
%---------
& &
3\,R_\kappa\,v_S
+\frac{9v^2}{4}\,{\cal R}\,s_{2\beta}
%---------
\end{array} \right) \ + \ M_A^2\cdot {\cal O}(\epsilon^4)\,, \nonumber
\end{eqnarray}
which gives rise to the eigenmasses
\begin{eqnarray}
M_{L_1,L_2}^2 &\approx & \frac{1}{2}\Bigg\{
\left[M_Z^2+v_S^2\,\left(2|\kappa|^2-\frac{R_\kappa}{v_S}\right)
-\frac{3}{4}v^2s_{2\beta}{\cal R}\right]
\nonumber \\ && \hspace{0.5cm}
\pm
\sqrt{\left[M_Z^2-v_S^2\,\left(2|\kappa|^2-\frac{R_\kappa}{v_S}\right)
+\frac{3}{4}v^2s_{2\beta}{\cal R}\right]^2
+ \left[\frac{v}{v_S}\,Y+vv_Ss_{2\beta}{\cal R}\right]^2 
} \Bigg\}\,, ~~~~
\label{eq:ml12} \\[0.3cm]
M_{L_3}^2 &\approx &
3R_\kappa v_S + \frac{9}{4} v^2 s_{2\beta}{\cal R}\,.
\label{eq:ml3}
\end{eqnarray}
We note that the CP-mixing entries in
the heavier-state matrix $\widetilde{{\cal M}_J^2}$ and 
the lighter matrix $\widetilde{{\cal M}_L^2}$
are proportional to the factors $v^2\,{\cal I}\,s_{2\beta}$ or
$vv_S\,{\cal I}\,s_{2\beta}$, respectively, 
and would not affect the approximated mass spectrum
up to the order $O(\epsilon^2)$.
This could be easily understood by observing the CP-mixing entries
in ${\cal M}_{JL}^2$ (\ref{eq:MJL})
and ${\cal M}_{L}^2$ (\ref{eq:ML}), which are proportional to ${\cal I}$,
are suppressed by the factor $\epsilon^2$
and $\epsilon^3$, respectively.
In the CP-conserving limit or up to the order $O(\epsilon^2)$,
our results agree with those
in Ref.~\cite{Miller:2003ay}.

When the $U(1)$ PQ symmetry is not broken or 
$|\kappa|={\cal R}={\cal I}={R}_\kappa/v_S=0$, 
there is no CP-violating mixing
and the determinant of the lower-right 
$2 \times 2$ submatrix of Eq.~(\ref{eq:mNH})
for the CP-odd states vanishes,
resulting in a massless CP-odd PQ axion; or else 
its mass is approximately given by 
Eq.~(\ref{eq:ml3}).
In the same  PQ-symmetric limit,  the lighter CP-even state
becomes tachyonic unless $Y=0$ as seen from Eq.~(\ref{eq:ml12}).
When the $U(1)$ PQ symmetry is broken,
it is interesting to note that  the condition
$M_{L_1}^2\geq 0$ gives
\begin{equation}
\bigg|M_A^2s_{2\beta}^2 -2|\lambda|^2v_S^2+v_S^2s_{2\beta}{\cal R}
\bigg| \leq \frac{2v_S M_Z}{v}\,
\sqrt{2|\kappa|^2v_S^2-R_kv_S-\frac{3}{4}{\cal R}v^2s_{2\beta}}\,.
\end{equation}
This, together with Eq.~(\ref{eq:ma}),
leads to the following 
constraints on the parameter space in the leading order:
\begin{eqnarray}
&& 0 \lsim 
R_\kappa = \frac{|\kappa| |A_\kappa|}{\sqrt{2}}\,
\cos(\phi^\prime_\kappa+\phi_{A_\kappa})\,
\lsim 2\, v_S\, |\kappa|^2\,,
\nonumber \\
&& M_{H^\pm}^{(0)} \sim M_{J_1,J_2} \sim M_A \sim
\frac{\sqrt{2}\,|\lambda|\,v_S}{s_{2\beta}}\,,
\nonumber \\
&&|A_\lambda| \sim \frac{\sqrt{2}\,|\lambda|\,v_S}
{s_{2\beta}\,\cos(\phi^\prime_\lambda+\phi_{A_\lambda})}
\sim \frac{M_A}{\cos(\phi^\prime_\lambda+\phi_{A_\lambda})} \,,
\end{eqnarray}
which more or less 
lift up the twofold ambiguity in
using the two CP-odd tadpole conditions by fixing
\begin{equation}
{\rm sign}\,[\cos(\phi^\prime_\kappa+\phi_{A_\kappa})] =
{\rm sign}\,[\cos(\phi^\prime_\lambda+\phi_{A_\lambda})]  = + 1 \,.
\end{equation}

%-------------------
\subsection{Vacuum condition}
%-------------------

Considering neutral fields only,
the effective potential at tree level takes the form
\begin{eqnarray}
\widetilde{V_0}(\tilde{\boldsymbol{v}})
&=&\frac{1}{2}m_1^2\tilde v_d^2+\frac{1}{2}m_2^2\tilde v_u^2
        +\frac{1}{2}m_S^2\tilde v_S^2-R_\lambda \tilde v_d\tilde v_u\tilde
v_S-\frac{1}{3}R_\kappa \tilde v^3_S 
\nonumber \\ && 
        +\frac{g_2^2+g_1^2}{32}(\tilde v_d^2-\tilde v_u^2)^2
        +\frac{|\lambda|^2}{4}(\tilde v_d^2\tilde v_u^2+\tilde v_d^2\tilde v_S^2
        +\tilde v_u^2\tilde v_S^2)
\nonumber \\ && 
        +\frac{|\kappa|^2}{4}\tilde v^4_S
        -\frac{1}{2}\mathcal{R}\tilde v_d\tilde v_u\tilde v^2_S\,,
\label{eq:vac}
\end{eqnarray}
where $\tilde{\boldsymbol{v}}=(\tilde{v}_u, \tilde{v}_d, \tilde{v}_S, 
\tilde{\theta}, \tilde{\varphi})$ denotes the arbitrary constant fields.
The CP phases of the VEVs enter into the potential
through ${\cal R}$, $R_\lambda$, and $R_\kappa$ under
the CP-odd tadpole conditions.
In the NMSSM, diverse vacua can exist. 
Following the classification discussed in Ref.~\cite{Funakubo:2005pu}, we 
define the following phases:
\begin{eqnarray}
{\rm EW} \!\!\!&:&\!\!\! v\neq0,~v_S\neq0,\nonumber \\
{\rm I}\!\!\!&:&\!\!\! v=0,~v_S\neq0,\nonumber \\
{\rm II}\!\!\!&:&\!\!\! v\neq0,~v_S=0,\nonumber \\
{\rm SYM}\!\!\!&:&\!\!\! v=v_S=0,
\label{eq:vevs}
\end{eqnarray} 
where $v=\sqrt{v^2_d+v^2_u}$.
Unlike the MSSM at the tree level, the electroweak-broken vacuum (denoted by EW) 
is not necessarily the global minimum
due to the presence of the cubic terms in the Higgs potential
which are proportional to $R_\lambda$ and $R_\kappa$.
To ensure the presumed vacuum is the true global minimum of the potential, 
we require the vacuum energy 
for the chosen vacuum to be smaller than that for any other choices
\begin{equation}
V_0^{\rm EW}\equiv
\widetilde{V_0}(\tilde{\boldsymbol{v}}=\boldsymbol{v}) <
\widetilde{V_0}(\tilde{\boldsymbol{v}}\neq\boldsymbol{v}).
\label{eq:VAC_COND}
\end{equation}
After removing the soft masses $m_{1,2,S}^2$ using the CP-even tadpole conditions,
the energy level of the EW vacuum can be rewritten as
\begin{eqnarray}
V_0^{\rm EW} &= & \frac{1}{8}\,v^2\,\left\{ s_{2\beta}^2\,\left[
\left(\frac{2v_SR_\lambda+v_S^2{\cal R}}{s_{2\beta}}\right)-\frac{|\lambda|^2v^2}{2}
\right] -2\,|\lambda|^2\,v_S^2 -M_Z^2\,c_{2\beta}^2\right\}
\nonumber \\[0.2cm] 
&+&\frac{\cal R}{8}\,v^2v_S^2\,s_{2\beta}
-\frac{|\kappa|^2}{4}\,v_S^4+\frac{R_\kappa}{6}\,v_S^3\,,
\label{eq:EW_vac}
\end{eqnarray}
where $c_\beta=\cos\beta=v_d/v$ and
$s_\beta=\sin\beta=v_u/v$.
Here, as an example, we compare the vacuum energy of the phase EW 
with that of the phase II.
The energy difference, which
should be positive definite if
the EW vacuum is the global minimum, is given by
\begin{eqnarray}
\Delta V_0^{{\rm II}-{\rm EW}}
\!\!\!&\equiv&\!\!\!
V_0^{\rm II}-V_0^{\rm EW} \nonumber\\
\!\!\!&=&\!\!\!	-\frac{1}{2}R_\lambda v_dv_uv_S-\frac{1}{6}R_\kappa v^3_S
	-\frac{1}{2}\mathcal{R}v_dv_uv^2_S
	+\frac{g^2_2+g^2_1}{32}\Big\{(\bar{v}^2_d-\bar{v}^2_u)^2-(v^2_d-v^2_u)^2\Big\}
	\nonumber\\
\!\!\!&&\!\!\!	-\frac{|\lambda|^2}{4}\Big\{(\bar{v}^2_d\bar{v}^2_u-v^2_dv^2_u)
	-v^2_dv^2_S-v^2_uv^2_S\Big\}+\frac{|\kappa|^2}{4}v^4_S,
\label{eq:II-EW_vac}
\end{eqnarray}
where $V_0^{\rm II}$ denotes the vacuum energy of the phase II
with VEVs $\bar{\boldsymbol{v}}\equiv
(\bar{v}_d\,,\bar{v}_u\,,0\,,0\,,0)$. 
Since the two vacua $\bar{\boldsymbol{v}}$ and $\boldsymbol{v}$ 
are the simultaneous solutions to the same tadpole conditions,
the VEVs $\bar{v}_d$ and $\bar{v}_u$ 
are determined by the tadpole conditions
taking $v_S=0$ and using the same soft masses 
$m_{1,2,S}^2$ fixed by the EW vacuum.
It should be noted that $\Delta V_0^{{\rm II}-{\rm EW}}$ can be negative 
if $R_\lambda$ and/or $R_\kappa$ with the positive signs become large,
leading to a metastable EW vacuum. 
Comparison of the EW vacuum with the other vacua also gives
similar expressions as Eq.~(\ref{eq:II-EW_vac}).
Therefore, the requirement for the global minimum can constrain the size of 
$|A_\lambda|$ and $|A_\kappa|$
\footnote{
Although a sufficiently long-lived metastable vacuum may be viable,
we will not consider such a case in this work.
}.

%%%%%%%%%%%%%%%%%%%%%-------------------
\setcounter{equation}{0}
\section{Higgs sector at the one-loop level}
%%%%%%%%%%%%%%%%%%%%%-------------------
The one-loop contributions to the Higgs-boson masses can be computed from 
the effective potential~\cite{Coleman:1973jx,Jackiw:1974cv}
\begin{equation}
V_1 =\frac{1}{64\pi^2}\,{\rm Str}\,\left[
{\cal M}^4\left(\log\frac{{\cal M}^2}{Q_0^2}-\frac{3}{2}\right)
\right]
\end{equation}
where $Q_0$ is the renormalization scale and
${\cal M}$ is the field-dependent mass matrix of all modes that couple 
to the Higgs bosons. 
The supertrace is defined as Str$[f({\cal M}^2)] \equiv \sum_i C_i (-1)^{2s_i}
(2 s_i  + 1)[f(m_i^2)]$, where $C_i$ is the color degrees of freedom
and $s_i$ is the spin of the $i^{\rm th}$ particle.
The field-dependent third-generation quark masses are given by
\begin{equation}
m_b^2 = |h_b|^2 | H_d^0 |^2\,; \ \ \
m_t^2 = |h_t|^2 | H_u^0 |^2\,,
\end{equation}
where $H^0_{d,u}$ are the neutral components of $H_{d,u}$.
The corresponding eigenvalues of the squark mass matrices are
\begin{eqnarray}
m_{\widetilde{t}_{1,2}}^{2}&=&
        \frac{1}{2}\bigg[M_{\widetilde{Q}}^{2}+M_{\widetilde{U}}^{2}
        +2|h_{t}|^{2}|H^0_{u}|^{2}+\frac{g_{2}^{2}
        +g_{1}^{2}}{4}\Big(|H^0_{d}|^{2}-|H^0_{u}|^{2}\Big) \nonumber \\
        & &\pm\sqrt{\Big[M_{\widetilde{Q}}^{2}-M_{\widetilde{U}}^{2}+x_{t}(|H^0_{d}|^{2}
        -|H^0_{u}|^{2})\Big]^{2}
        +4|h_{t}|^{2}\left|A_{t}H_u^0-\lambda^* S^* (H_d^{0})^*\right|^{2}}\bigg],
\nonumber \\
m_{\widetilde{b}_{1,2}}^{2}&=&
        \frac{1}{2}\bigg[M_{\widetilde{Q}}^{2}+M_{\widetilde{D}}^{2}
        +2|h_{b}|^{2}|H^0_{d}|^{2}-\frac{g_{2}^{2}
        +g_{1}^{2}}{4}\Big(|H^0_{d}|^{2}-|H^0_{u}|^{2}\Big) \nonumber \\
        & &\pm\sqrt{\Big[M_{\widetilde{Q}}^{2}-M_{\widetilde{D}}^{2}-x_{b}(|H^0_{d}|^{2}-
        |H^0_{u}|^{2})\Big]^{2}
        +4|h_{b}|^{2}|A_{b}H_d^0-\lambda^* S^*(H_u^0)^*|^{2}}\bigg],
\label{eq:msq}
\end{eqnarray}
where
\begin{equation}
x_{t}=\frac{1}{4}\bigg(g_{2}^{2}-\frac{5}{3}g_{1}^{2}\bigg)
\qquad {\rm and} \qquad
x_{b}=\frac{1}{4}\bigg(g_{2}^{2}-\frac{1}{3}g_{1}^{2}\bigg).
\end{equation}
In Eq.~(\ref{eq:msq}), $M_{\widetilde{Q}}^{2}$, $M_{\widetilde{U}}^{2}$,
and $M_{\widetilde{D}}^{2}$ are real soft-SUSY breaking parameters and
$A_t$ and $A_b$ are complex soft-SUSY breaking parameters, 
$A_q=|A_q|\,{\rm e}^{i\,\phi_{A_q}}$ with $q=t,b$.
For the explicit expressions of the corrections to the Higgs-boson mass matrix, 
we refer to Ref.~\cite{Funakubo:2004ka}
\footnote{In addition to the 
dominant contribution from third-generation quarks and squarks,
we have included contributions from 
the weak gauge bosons in our numerical analysis.}.

While the relation between $I_\kappa$ and ${\cal I}$
remains the same as in Eq.~(\ref{eq:ilik}),
the tree-level relation between $I_\lambda$ and ${\cal I}$ is modified
in the presence of the one-loop corrections as
\begin{equation}
I_\lambda = \frac{|\lambda| |A_\lambda|}{\sqrt{2}}\,
\sin(\phi^\prime_\lambda+\phi_{A_\lambda}) = - \frac{1}{2}\,\mathcal{I}\, v_S
 \ - \ \Delta I_\lambda \,,
\label{eq:il}
\end{equation}
where
\begin{equation}
\Delta I_\lambda = \frac{3}{16 \pi^2} \sum_{q=t,b}
|h_q|^2\, I_q \, 
f\left(\langle m^2_{\tilde{q}_1}\rangle,\langle m^2_{\tilde{q}_2}\rangle\right)\, 
\end{equation}
with 
\begin{equation}
I_q =\frac{|\lambda|\,|A_q|}{\sqrt{2}}\,\sin(\phi_\lambda^\prime+\phi_{A_q})
\end{equation}
and
\begin{equation}
f(a,b)=\frac{1}{a-b}\,\left[
a\left(\log\frac{a}{Q_0^2}-1\right)
-b\left(\log\frac{b}{Q_0^2}-1\right) \right]\,.
\end{equation}
The one-loop corrected charged Higgs-boson mass is given by
\begin{equation}
M_{H^\pm}^{2} = M_{H^\pm}^{(0)\,2} +\Delta M_{H^\pm}^{2}
\end{equation}
where we refer to Ref.~\cite{Funakubo:2002yb}
 for the explicit form of the correction $\Delta M_{H^\pm}^{2}$.

In this work, we have included logarithmically enhanced two-loop corrections
of the order ${\cal O}(g_s^2 h^4)$ and ${\cal O}(h^6)$, where 
$g_s$ is the strong gauge coupling.
We have adopted the algorithm suggested in Ref.~\cite{Haber:1996fp},
which incorporates the effects
of the RG improvement and minimizes the 
two-loop corrections.
Including the full one-loop radiative corrections, the $5\times 5$ 
mass matrix of the
neutral Higgs boson can be denoted by
\begin{equation}
{\cal M}^{2}_N= {\cal M}^{(0)\,2}_N + \Delta {\cal M}^{2}_N
\end{equation}
where the one-loop correction part may further be decomposed into
\begin{equation}
\Delta {\cal M}^{2}_N = 
\left(\Delta {\cal M}^{2}_{N}\right)_{\rm LL} + 
\left(\Delta {\cal M}^{2}_{N}\right)_{\rm mix}\,.
\end{equation}
The first term contains the genuine logarithmic contributions which 
are present even when
the left-right mixing of the third-generation squarks is absent.
The second term describes the threshold effects arising from the
mass splittings due to
the left-right mixing in the third-generation squark sectors.
The RG-improved mass matrix can be well approximated by
replacing $m_t$ and $m_b$ in each term 
by the running masses at appropriate scales as in
\begin{eqnarray}
\left(\Delta {\cal M}^{2}_N\right)_{\rm RG} &\simeq &
\overline{\left(\Delta {\cal M}^{2}_{N}\right)}_{\rm LL} + 
\overline{\left(\Delta {\cal M}^{2}_{N}\right)}_{\rm mix}\nonumber \\
& \equiv &
{\left(\Delta {\cal M}^{2}_{N}\right)}_{\rm LL}[m_t(\mu_t)\,,m_b(\mu_b)] +
{\left(\Delta {\cal M}^{2}_{N}\right)}_{\rm mix}
[m_t(\mu_{\tilde t})\,,m_b(\mu_{\tilde b})]
\end{eqnarray}
where the intermediate scales $\mu_t^2 = {m_t\,\mu_{\tilde t}}$ and 
$\mu_b^2 = {m_t\, \mu_{\tilde b}}$ with
\begin{eqnarray}
\mu_{\tilde t}^2 &=& \max\,(M_{\widetilde Q}^2+m_t^2, M_{\widetilde U}^2+m_t^2)\,, \nonumber \\
\mu_{\tilde b}^2 &=& \max\,(M_{\widetilde Q}^2+m_b^2, M_{\widetilde D}^2+m_b^2)\,.
\end{eqnarray}
To obtain the quark masses at the SUSY and intermediate scales, we used 
the RG equations (RGEs) of the two-Higgs doublet model
\begin{eqnarray}
\frac{{\rm d}\,m_b^2}{{\rm d}\ln\mu^2} & =&
\frac{1}{64\pi^2}\,\left[6 h_b^2 + 2 h_t^2 -32 g_s^2 \right]\,m_b^2 \,,
\nonumber \\
\frac{{\rm d}\,m_t^2}{{\rm d}\ln\mu^2} & =&
\frac{1}{64\pi^2}\,\left[6 h_t^2 + 2 h_b^2 -32 g_s^2 \right]\,m_t^2 \,,
\end{eqnarray}
when $\mu > M_A$
assuming $M_A < \max(\mu_{\tilde t},\mu_{\tilde b})$. 
When  $\mu \leq M_A$, the SM RGEs are used:
\begin{eqnarray}
\frac{{\rm d}\,m_b^2}{{\rm d}\ln\mu^2} & =&
\frac{1}{64\pi^2}\,\left[6 (h_b^{\rm SM})^2 - 6 (h_t^{\rm SM})^2 -32 g_s^2 \right]\,m_b^2 \,,
\nonumber \\
\frac{{\rm d}\,m_t^2}{{\rm d}\ln\mu^2} & =&
\frac{1}{64\pi^2}\,\left[6 (h_t^{\rm SM})^2 - 6 (h_b^{\rm SM})^2 -32 g_s^2 \right]\,m_t^2
\,.
\end{eqnarray}

%%%%%%%%%%%%%%%%%%%%%-------------------
\setcounter{equation}{0}
\section{Numerical analysis}
%%%%%%%%%%%%%%%%%%%%%-------------------
%
In this section, we
present the numerical results of the RG-improved calculation of the masses and
mixing matrix of the NMSSM Higgs bosons in the presence of 
nontrivial CP phases.
We make comparisons, where possible, with other RG-improved calculations 
without 
CP phases~\cite{nmhdecay} and the one-loop calculations 
including CP phases but without the RG improvement~\cite{Funakubo:2004ka}.
The input parameters for the Higgs sector in our numerical study are specified
as follows:
\begin{eqnarray}
{\rm tree~level} &:&  |\lambda|\,,|\kappa|\,,\tan\beta\,; \ |A_\lambda|\,,|A_\kappa|\,,v_S
\nonumber \\
\mbox{ 1-loop~level} &:&
M_{\widetilde Q}\,,M_{\widetilde U}\,,M_{\widetilde D}\,,
|A_t|\,,|A_b|
\nonumber \\
{\rm CP~phases} &:& \phi_\lambda^\prime\,,\phi_\kappa^\prime\ ; \ \phi_{A_t}\,,\phi_{A_b}
\nonumber \\
{\rm signs~of} &:& \cos(\phi_\lambda^\prime+\phi_{A_\lambda})\,,
\cos(\phi_\kappa^\prime+\phi_{A_\kappa})\,.
\end{eqnarray}
For the renormalization scale $Q_0$ we take the top-quark mass 
as in Refs.~\cite{Casas:1994us,CPmixing1,CPmixing2}.

On the scenarios under consideration in this section,
we have imposed the following three conditions:
\begin{itemize}
\item the LEP limits~\cite{Schael:2006cr},
\item the global minimum condition requiring that the electroweak vacuum
chosen by fixing the values of $\tan\beta$ and $v_S$
is the global minimum of the RG-improved effective potential, and 
\item the positiveness of the Higgs masses squared, $M_H^2>0$, abandoning the
parameter space in which one or more of the Higgs states become tachyonic.
\end{itemize}

In order to determine 
whether the prescribed EW vacuum is the global minimum or not,
we minimize the effective potential numerically.
First, the tadpole conditions are solved
for input VEVs $\boldsymbol{v}_{\rm in}$ and then the solution
is fed into the potential. Then the potential 
is numerically minimized in the range 
$(v_d\,,v_u |c_\theta|\,,v_u |s_\theta|\,,
v_S |c_\varphi|\,,v_S |s_\varphi|) < 10$ TeV using the 
downhill simplex (Nelder-Mead) method~\cite{ref:NR}. 
The output minimum is denoted by $\boldsymbol{v}_{\rm out}$.
In the acceptable cases, the numerically obtained
$\boldsymbol{v}_{\rm out}$ is exactly the 
same as the input $\boldsymbol{v}_{\rm in}$ up to the $Z_3$ symmetry.
Depending on the parameter space, however, 
some $\boldsymbol{v}_{\rm out}$ may exist such that 
$V_{\rm eff}(\boldsymbol{v}_{\rm out})< V_{\rm eff}(\boldsymbol{v}_{\rm in})$.
In most cases, $\boldsymbol{v}_{\rm out}
(\neq \boldsymbol{v}_{\rm in})$ corresponds to the VEVs in one of 
the three phases I, II, or SYM~(\ref{eq:vevs})
in which one or both $v$ and $v_S$ are vanishing.
When $\boldsymbol{v}_{\rm out}$ has $v_{u,d}\neq 0$ and $v_S\neq 0$ 
but with $v_{\rm out}\neq 246$ GeV,
it can be made acceptable in principle by rescaling the Higgs VEVs 
to make $v_{\rm out}=246$ GeV.
But the rescaling may require compensating changes 
of the original values of $\tan\beta$ and $v_S$ to satisfy the 
tadpole conditions.
In this work, we discard such cases by keeping $\tan\beta$ and $v_S$ fixed.

Before we go into various scenarios we offer a comment about the
range of $\tan\beta$. It has been shown in Ref.~\cite{Funakubo:2004ka} 
and we have also verified that large $\tan\beta > 20$~GeV is not favored by the
LEP and global minimum constraints.  Furthermore, the CP-violating
scalar-pseudoscalar mixing is suppressed by large $\tan\beta$. We
therefore employ a small to moderate $\tan\beta$ in the following
scenarios.
%-------------------
\subsection{A typical scenario}
%-------------------
%
We first consider a typical scenario in which the 
heavier Higgs bosons are relatively light by choosing
a small $\tan\beta$ and a moderate $v_S$~\cite{Miller:2003ay}:
\begin{eqnarray}
&&
\tan\beta=3\,, \ \
v_S=750~{\rm GeV}\,,
\nonumber \\
&&
M_{\widetilde Q}=M_{\widetilde U}=M_{\widetilde D}=|A_t|=|A_b|=1000~{\rm GeV}\,,
\nonumber \\
&&
\phi_\lambda^\prime=0\,, \ \
\phi_{A_t}=\phi_{A_b}=0\,,
\nonumber \\
&&
{\rm sign}\,[\cos(\phi^\prime_\kappa+\phi_{A_\kappa})] =
{\rm sign}\,[\cos(\phi^\prime_\lambda+\phi_{A_\lambda})]  = + 1 \,,
\label{eq:scn1}
\end{eqnarray}
while varying
\begin{equation}
|\lambda|\,,|\kappa|\,; \ \ 
|A_\lambda|\,,|A_\kappa| \,; \ \
\phi_\kappa^\prime\,.
\end{equation}
For definiteness we have fixed 
$M_1=M_2=-200$ GeV.

\begin{figure}[t!]
\begin{center}
\includegraphics[width=11cm]{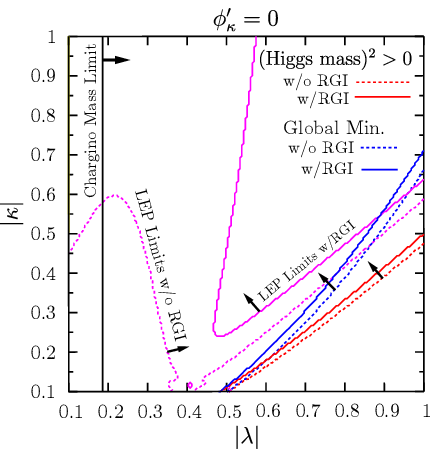}
\includegraphics[width=7cm]{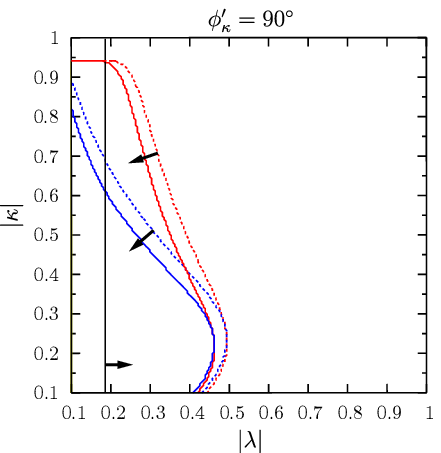}
\includegraphics[width=7cm]{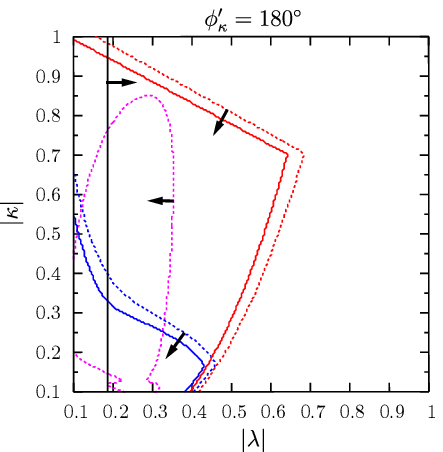}
\end{center}
\caption{The allowed region in the $|\kappa|$-$|\lambda|$ plane for
the scenario in Eq.~(\ref{eq:scn1}) with
$|A_\lambda|=500$ GeV and $|A_\kappa|=100$ GeV for three values of
$\phi^\prime_\kappa=0^\circ$ (upper panel),
$\phi^\prime_\kappa=90^\circ$ (lower left panel), and
$\phi^\prime_\kappa=180^\circ$ (lower right panel). 
RGI stands for the RG improvement.}
\label{fig:kap_lam_C1}
\end{figure}
In Fig.~\ref{fig:kap_lam_C1}, we show the allowed region 
satisfying the three conditions
in the $|\kappa|$-$|\lambda|$ plane when
$|A_\lambda|=500$ GeV and $|A_\kappa|=100$ GeV for 
three values of the tree-level CP phase:
$\phi^\prime_\kappa=0^\circ$ (upper panel),
$90^\circ$ (lower left panel), and $180^\circ$ (lower right panel).

The magenta, blue, and red lines denote the LEP limits, 
the global minimum condition, and the positiveness of the Higgs mass
squared, respectively.
The solid (dotted) lines
are after (before) the inclusion of the RG improvement.
The vertical solid line denotes the chargino mass limit, 
$m_{\tilde\chi^\pm}>104$ GeV~\cite{Abbiendi:2003sc}.
For $\phi^\prime_\kappa=0^\circ$,
we observe that the LEP constraint is significantly 
strengthened by the RG improvement, allowing only relatively large
couplings with
$|\lambda|\gsim 0.45$ and $|\kappa|\gsim 0.22$. In this regard,
the parameter space point of $|\lambda|=0.3$ and $|\kappa|=0.1$ considered in
Ref.~\cite{Miller:2003ay}, which is marginally compatible with
the LEP limits without the RG improvement, is completely ruled out
with the inclusion of the RG improvement.
For the condition of $M_H^2>0$, we see that too large 
$|\lambda|$ is not allowed when $|\kappa| \lsim 0.5$.
This can be understood from Eq.~(\ref{eq:ml12}) by observing that
due to the third negative term in the first line and
the second term inside the square root below,
the lighter state mass
$M_{L_1}^2$ could become negative when $|\lambda|$ is too large,
unless the term $2|\kappa|^2v_S^2$ in the first line is large enough.
Nevertheless, after imposing the current LEP bounds, we see that the condition
does not rule out more regions in the $|\kappa|$-$|\lambda|$ plane.
On the other hand,
we see that the global minimum condition is always stronger than
the $M_H^2>0$ condition and 
it is more restrictive than
the LEP bounds in the region $0.55 \lsim |\kappa| \lsim 0.7$ 
when $|\lambda|\gsim 0.9$.
For $\phi^\prime_\kappa=90^\circ$, the 
theoretically allowed region is 
significantly reduced and no region remains after applying
the LEP limits.
For $\phi^\prime_\kappa=180^\circ$, 
we still see no region compatible with the LEP bounds
after the inclusion of the RG improvement.

\begin{figure}[t!]
\begin{center}
\includegraphics[width=5.2cm]{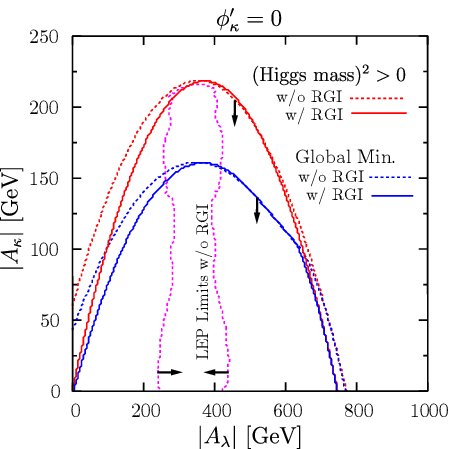}
%\hspace{0.5cm}
\includegraphics[width=5.2cm]{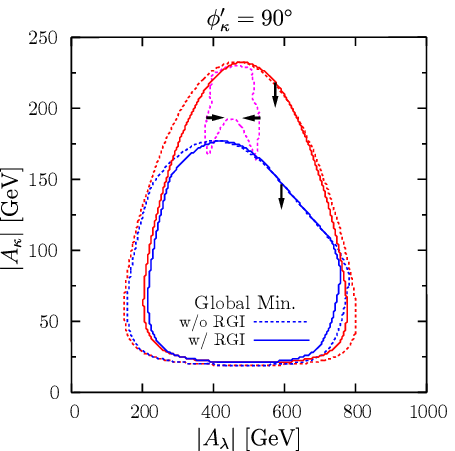}
%\\[0.5cm]
\includegraphics[width=5.2cm]{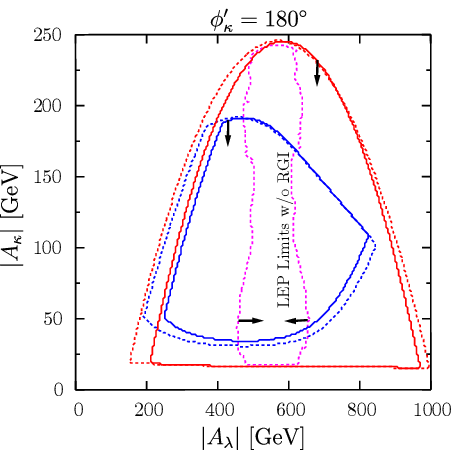}
\end{center}
\caption{The allowed region in the $|A_\kappa|$-$|A_\lambda|$ plane 
for the three values of 
$\phi^\prime_\kappa=0^\circ$ (left panel),
$90^\circ$ (middle panel), and $180^\circ$ (right panel).
The scenario in Eq.~(\ref{eq:scn1}) is taken with
$|\lambda|=0.3$ and $|\kappa|=0.1$. 
The lines are the same as in Fig.~\ref{fig:kap_lam_C1}.}
\label{fig:Akap_Alam_C1}
\end{figure}
In Fig.~\ref{fig:Akap_Alam_C1}, we show the allowed region 
in the $|A_\kappa|$-$|A_\lambda|$ plane for the three values of
$\phi^\prime_\kappa=0^\circ\,,90^\circ\,,$ and $180^\circ$. We have taken
$|\lambda|=0.3$ and $|\kappa|=0.1$ for comparisons with
the results presented in \cite{Miller:2003ay}, though
this point has been ruled out by the LEP limits taking into
account the RG improvement. Note that we only show the LEP limits without the
RG improvement here.
By changing $\phi^\prime_\kappa$ from $0^\circ$ to $180^\circ$, we observe that
the
allowed region moves to the direction of increasing $|A_\lambda|$
and $|A_\kappa|$. Specifically, $|A_{\kappa\,,\lambda}|=0$ GeV is not allowed 
when $\phi^\prime_\kappa>0^\circ$
due to the CP-odd tadpole conditions with
${\cal I}\neq 0$; see Eq.~(\ref{eq:ilik}).
We also observe that the allowed region is very small when CP is maximally
violated with 
$\phi^\prime_\kappa=90^\circ$; see the middle frame.
This implies that
the inclusion of the CP phases may change the phenomenological features
of the Higgs sector significantly together with the RG improvement. 
We pursue this issue further in the next section with
a scenario compatible with the LEP limits.
Otherwise, a similar discussion can be applied as in
Fig.~\ref{fig:kap_lam_C1}: $(i)$ the global minimum condition is
always stronger than the $M_H^2>0$ condition and $(ii)$ it further 
constrains the parameter space in addition to the LEP limits.

\begin{figure}[t!]
\begin{center}
\includegraphics[width=5cm]{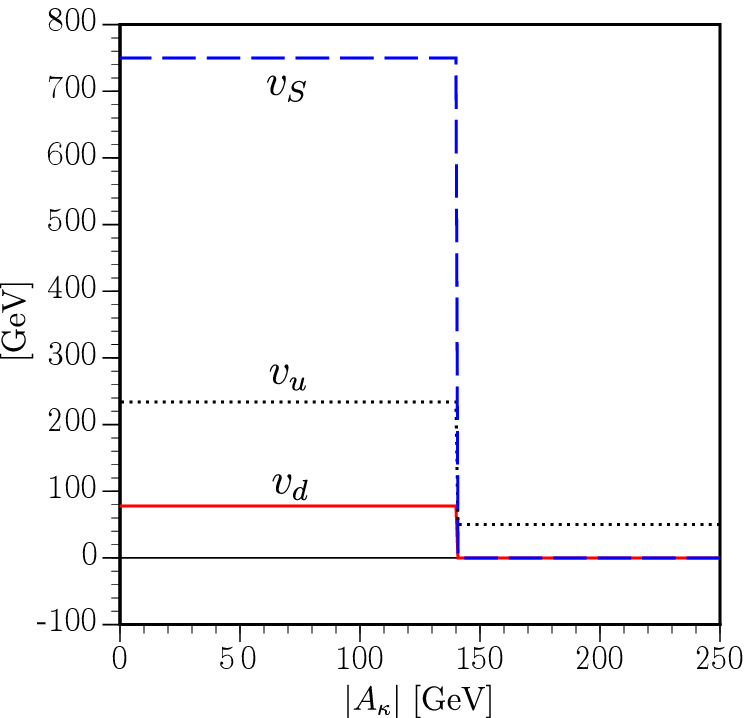}
%\hspace{0.5cm}
\includegraphics[width=5.3cm]{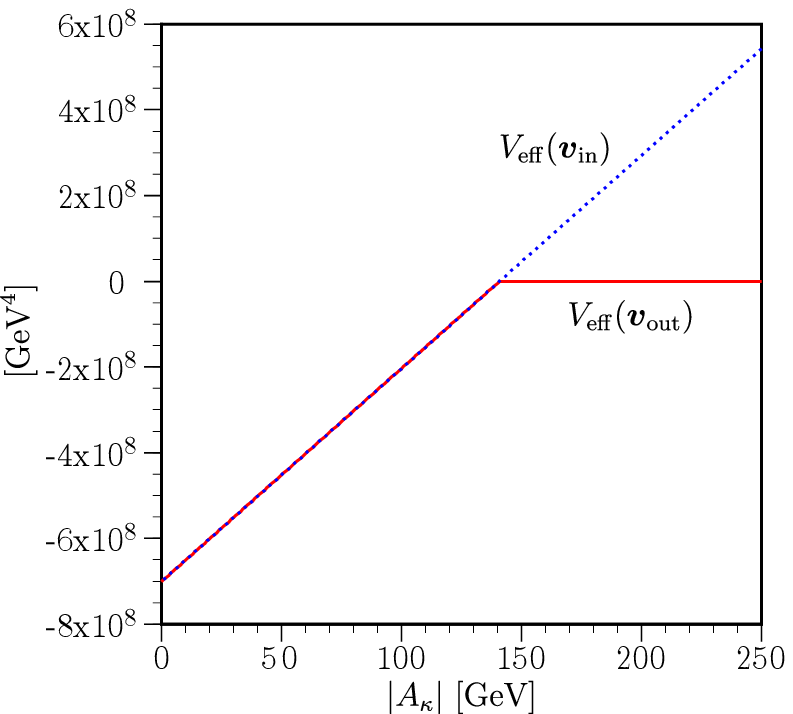}
%\\[0.5cm]
\includegraphics[width=5.3cm]{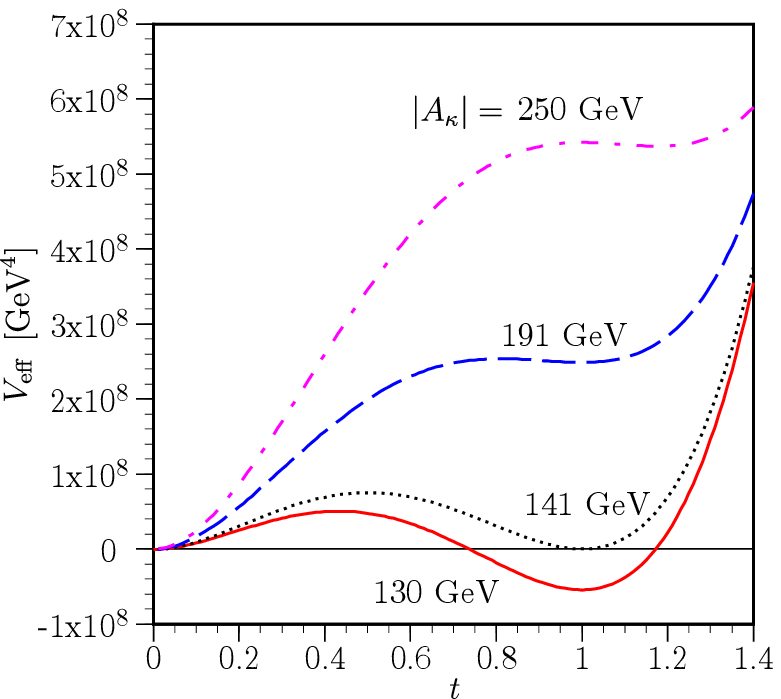}
\end{center}
\caption{ The output VEVs $v_u$, $v_d$, and $v_S$ as functions of
$|A_\kappa|$ (left panel) for the scenario in Eq.~(\ref{eq:scn1}) 
taking $|\lambda|=0.3$, $|\kappa|=0.1$, $|A_\lambda|=500$ GeV, and
$\phi^\prime_\kappa=0$.
The middle panel shows the vacuum energies for the input
and output vacua as functions of
$|A_\kappa|$, which
separates from each other when $|A_\kappa|$ is larger than 141 GeV.
In the right panel,
the vacuum energies along the direction 
connecting the II ($t=0$) and EW ($t=1$)
vacua are shown for several values of $|A_\kappa|$.
}
\label{fig:metavacua}
\end{figure}
The left plot in Fig.~\ref{fig:metavacua} shows the numerically obtained 
VEVs ($\boldsymbol{v}_{\rm out}$) as a function of $|A_\kappa|$
taking $\phi^\prime_\kappa=0$ and $|A_\lambda|=500$ GeV
with $|\lambda|=0.3$ and $|\kappa|=0.1$.
In the region, $0\le|A_\kappa|\lsim 141$ GeV, the prescribed VEVs 
$\boldsymbol{v}_{\rm EW}=\boldsymbol{v}_{\rm in}$=($v_d=78$ GeV, $v_u=234$ GeV, $v_S=750$ GeV)  agree with $\boldsymbol{v}_{\rm out}$ and thus satisfy the global minimum condition. 
For $|A_\kappa|\gsim 141$ GeV, however, the global minimum moves 
to phase II with 
$\boldsymbol{v}_{\rm II}=$($v_d=0$ GeV, $v_u=50$ GeV, $v_S=0$ GeV).
The energy levels of the two vacua are plotted in the middle panel.
The EW vacuum energy ($V_{\rm eff}(\boldsymbol{v}_{\rm in})$) becomes larger
as $|A_\kappa|$ increases. 
The linear dependence  is coined from the $R_\kappa$ term 
appearing in the potential at  tree level, see Eq.~(\ref{eq:EW_vac}),
with no $|A_\kappa|$-dependent terms in the one-loop effective potential.
On the other hand, the vacuum energy of phase II
$V^{\rm II}_0=-(g^2_2+g^2_1)v^4_u/32$+(one-loop corrections) 
$\sim -3\times 10^5$ GeV$^4$
is independent of $|A_\kappa|$ as shown in the figure.
In the right panel, for a given value of $|A_\kappa|$
the effective potential is plotted along the 
direction  which connects the two vacua
$\boldsymbol{v}_{\rm II}=(v_d=0$ GeV, $v_u=50$ GeV, $v_S=0$ GeV)
and $\boldsymbol{v}_{\rm EW}=(v_d=78$ GeV, $v_u=234$ GeV, $v_S=750$ GeV). 
The direction is parametrized by $t$ such that
\begin{eqnarray}
\boldsymbol{v}=(\boldsymbol{v}_{\rm EW}-\boldsymbol{v}_{\rm II})t+\boldsymbol{v}_{\rm II}\,,
\end{eqnarray}
and we are moving from the first minimum ($\boldsymbol{v}_{\rm II}$)
to the second extremum ($\boldsymbol{v}_{\rm EW}$)
as the parameter $t$ increases from $0$ to $1$.
We note that the EW vacuum is the global minimum
when $|A_\kappa|=130$ GeV (or $|A_\kappa|<141$ GeV)
and it becomes degenerate with the phase II vacuum when 
$|A_\kappa|\simeq 141$ GeV. If $|A_\kappa|$ increases further,
the EW vacuum is destabilized or $M_H^2<0$ when
$|A_\kappa|\simeq191$ GeV and beyond the point, it turns into the 
maximum as shown by the line with $|A_\kappa|=250$ GeV.

\begin{figure}[t!]
\begin{center}
\includegraphics[width=7.3cm]{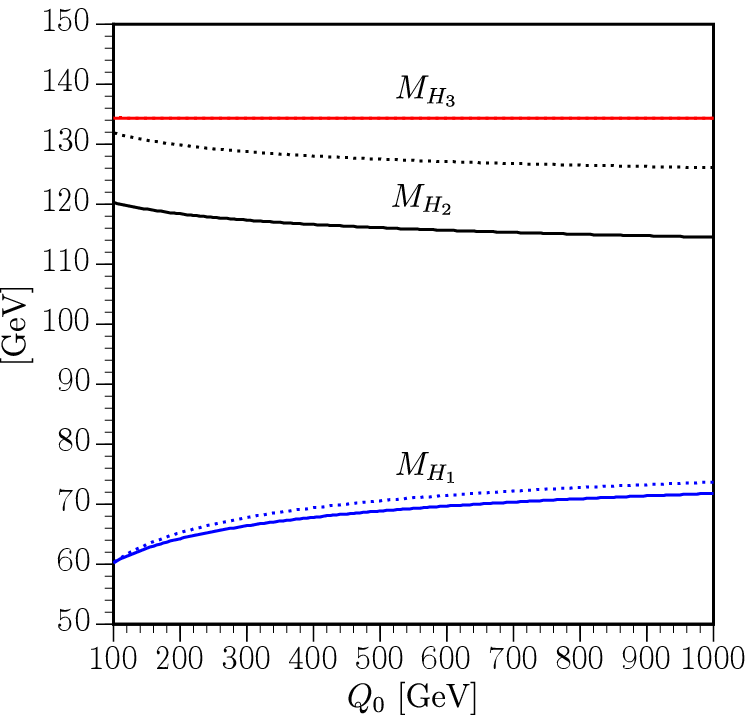}
%\hspace{0.5cm}
\includegraphics[width=7.3cm]{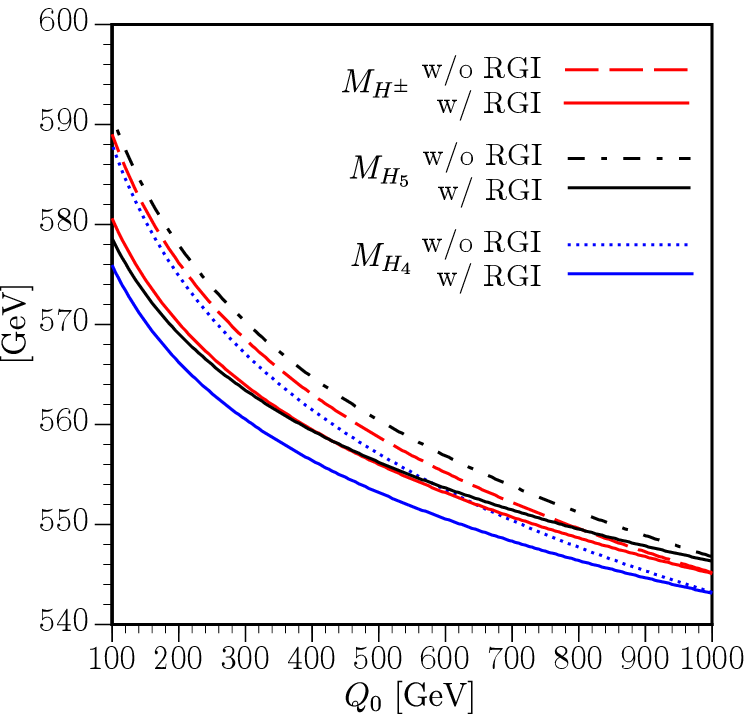}
%\\[0.5cm]
\end{center}
\caption{ The light (left panel) and heavy (right panel) Higgs boson masses as functions of
the renormalization scale $Q_0$ for the scenario in Eq.~(\ref{eq:scn1})
taking $|\lambda|=0.3$, $|\kappa|=0.1$, $|A_\lambda|=500$ GeV, 
$|A_\kappa|=100$ GeV, and $\phi^\prime_\kappa=0$,
with (solid lines) and without (dashed lines) RGI.
}
\label{fig:q0}
\end{figure}
Before moving to the next scenario, we examine the
renormalization scale ($Q_0$) dependence of the Higgs-mass spectrum.
In Fig.~\ref{fig:q0}, we show the masses of the lower three (left panel)
and the heavier two (right panel) neutral Higgs states in the range
between $Q_0=100$ GeV to $1000$ GeV. We consider the 
CP-conserving case with $\phi^\prime_\kappa=0$ taking
$|\lambda|=0.3$, $|\kappa|=0.1$, $|A_\lambda|=500$ GeV, and
$|A_\kappa|=100$ GeV.
First of all, we observe that the implementation of the 
RG improvement decreases the $Q_0$ dependence of the masses.
The heavier states show the larger variation in their masses
than the lighter ones.
Among the lighter states
the singlet CP-odd state $H_3$ is hardly affected by the loop corrections
due the small value of $|\kappa|=0.1$
and the $H_1$ state is less affected by the RG improvement than
$H_2$ but shows stronger dependence on $Q_0$.
We find similar behaviors in the results obtained by using 
the {\tt NMHDECAY} code~\cite{nmhdecay}.
In the CP-conserving limit,
the {\tt NMHDECAY} results are in agreement with ours after
taking into account the $Q_0$ dependence and
subleading contributions ignored in our calculation.
%
%-------------------
\subsection{A LEP-compatible scenario}
%-------------------
%
\begin{figure}[t!]
\begin{center}
\includegraphics[width=5.2cm]{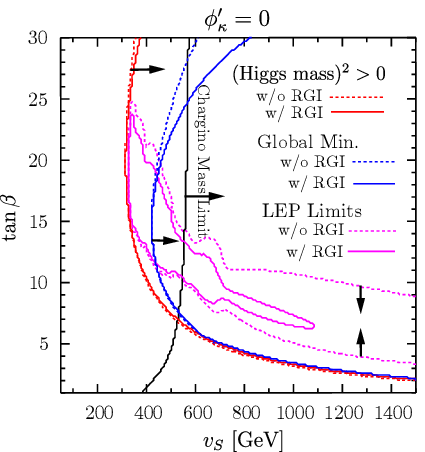}
%\hspace{0.5cm}
\includegraphics[width=5.2cm]{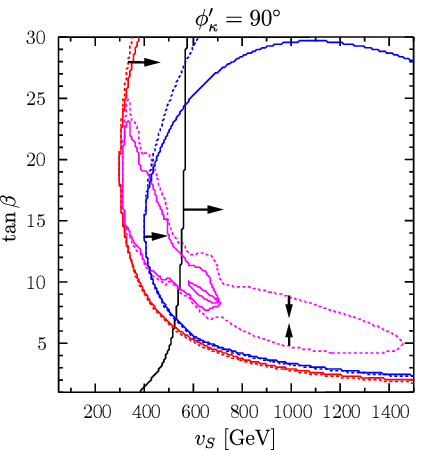}
%\\[0.5cm]
\includegraphics[width=5.2cm]{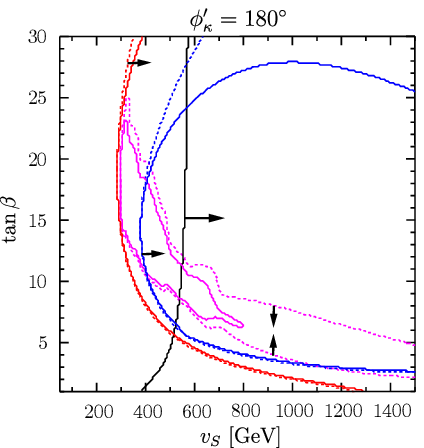}
\end{center}
\caption{
The allowed region in the $\tan\beta$-$v_S$ plane
for the three values of
$\phi^\prime_\kappa=0^\circ$ (left panel),
$90^\circ$ (middle panel), and $180^\circ$ (right panel).
We have taken
$|\lambda|=0.3$, $|\kappa|=0.15$, 
$|A_\lambda|=1200$ GeV, and $|A_\kappa|=130$ GeV
with 
$M_{\widetilde Q}=M_{\widetilde U}=M_{\widetilde D}=|A_t|=|A_b|=1000~{\rm GeV}$,
$\phi_\lambda^\prime=\phi_{A_t}=\phi_{A_b}=0$,
$M_1=M_2=-200$ GeV, and
${\rm sign}\,[\cos(\phi^\prime_\kappa+\phi_{A_\kappa})] =
{\rm sign}\,[\cos(\phi^\prime_\lambda+\phi_{A_\lambda})]  = + 1$.
}
\label{fig:C2.1}
\end{figure}
\begin{figure}[t!]
\begin{center}
\includegraphics[width=7.67cm]{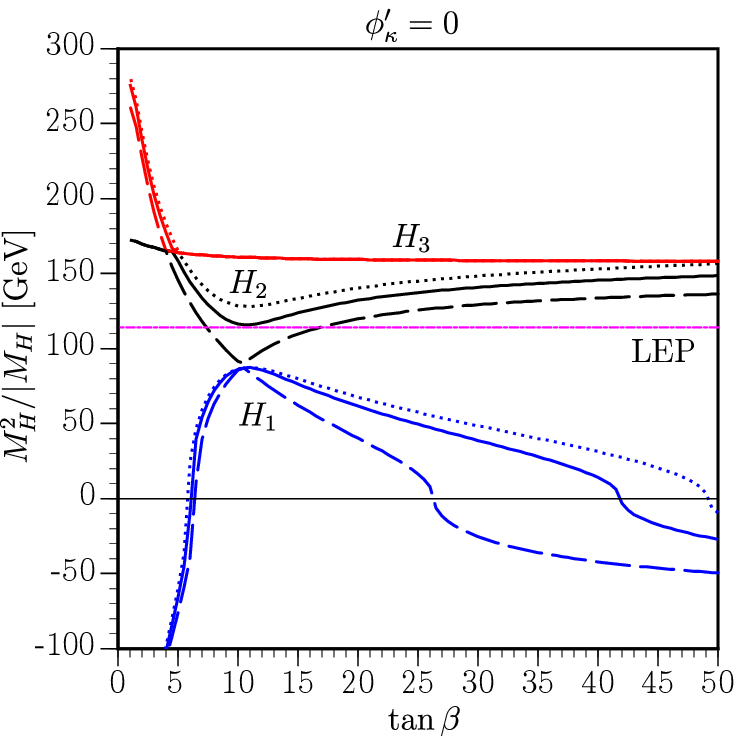}
\includegraphics[width=7.3cm]{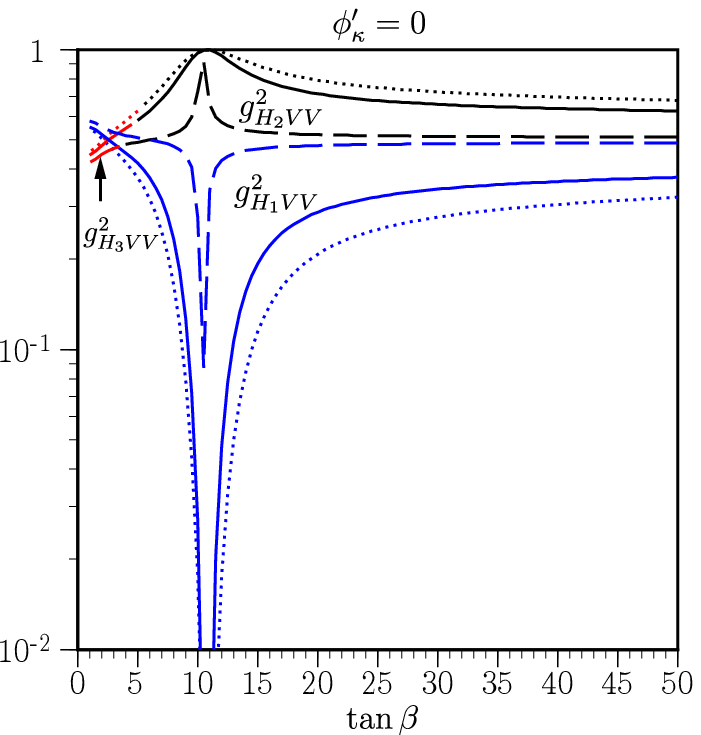}
%\includegraphics[width=5.2cm]{mH_tb_delkap0_C2.eps}
%%\hspace{0.5cm}
%\includegraphics[width=5.2cm]{gHVVsq_tb_delkap0_C2.eps}
%%\\[0.5cm]
%\includegraphics[width=5.2cm]{Osq_tb_delkap0_C2.eps}
\end{center}
\caption{
The masses $M_{H_i}^2/|M_{H_i}|$ (left panel) and couplings $g^2_{H_iVV}$ (right panel) 
as functions of $\tan\beta$.
The dashed lines are at the tree level and the
dotted and solid lines
at the one-loop level without and with the RGI, respectively.
We fix $v_S=600$ GeV while other parameters are 
the same as in Fig.~\ref{fig:C2.1}.
}
\label{fig:C2.2}
\end{figure}
The scenario considered in the previous section is not compatible with
the LEP limits after taking into account  
the RG improvement. Moreover, we observe that
the LEP limits become very strong with nontrivial CP phases.
In this section, we move to
a higher value  of $\tan\beta$ toward a LEP-compatible scenario independently
of CP phases. 

In Fig.~\ref{fig:C2.1}, we show
the allowed region in the $\tan\beta$-$v_S$ plane
for the three values of
$\phi^\prime_\kappa=0^\circ$ (left panel),
$90^\circ$ (middle panel), and $180^\circ$ (right panel).
We observe that $\tan\beta$ is bounded above by $\sim 18$
due to the LEP and global minimum constraints
and further we have $\tan\beta \lsim 13$ by applying the chargino mass limit.
We note the upper limit
on $\tan\beta$ is almost independent of $\phi'_\kappa$.
In Fig.~\ref{fig:C2.2} we show 
the masses $M_{H_i}^2/|M_{H_i}|$ (left panel) and couplings $g^2_{H_iVV}$ (right panel) 
for the lighter three states
as functions of $\tan\beta$ while fixing $v_S=600$ GeV.
The mass of $H_2$ is always above the SM LEP limit while
$M_{H_1}$ is below it.
For $\tan\beta \gsim 5$ where $M_{H_1}^2>0$, we find that $H_3 \sim a_S$ and 
$H_1$ and $H_2$ are mixtures mostly of $\phi_u$ and $\phi_S$
\footnote{
The heavier neutral states are such as $H_4 \sim a$ and $H_5 \sim \phi_d$.}.
We find that the mass difference between the two lightest states
becomes the smallest when $\tan\beta \sim 10$ and $H_1$ becomes dominated by the
singlet component there, explaining why the coupling $g_{H_1VV}$ 
almost vanishes; see the right frame. 
The $\tan\beta$ value at which the resonance occurs could be inferred
from Eq.~(\ref{eq:ml12}) by requiring $Y = 0$ 
\footnote{
We note $M_Z^2\simeq v_S^2\,(2|\kappa|^2-R_\kappa/v_S)$ for 
the parameters chosen.}
or $\tan\beta \sim \sqrt{2}|A_\lambda|/|\lambda|v_S \sim 10$.
As $\tan\beta$ grows, $M_{H_1}$ decreases and the size of the
coupling $g_{H_1VV}$ increases, leading us to the fact that
the large value of $\tan\beta$ is not compatible with
the LEP limits.

Combining these observations, we have fixed 
the parameters of our LEP-compatible scenario as

\begin{eqnarray}
&&
\tan\beta=10\,, \ \
v_S=600~{\rm GeV}\,,
\nonumber \\
&&
M_{\widetilde Q}=M_{\widetilde U}=M_{\widetilde D}=|A_t|=|A_b|=1000~{\rm GeV}\,,
\nonumber \\
&&
\phi_\lambda^\prime=0\,, \ \
\phi_{A_t}=\phi_{A_b}=0\,,
\nonumber \\
&&
{\rm sign}\,[\cos(\phi^\prime_\kappa+\phi_{A_\kappa})] =
{\rm sign}\,[\cos(\phi^\prime_\lambda+\phi_{A_\lambda})]  = + 1 \,,
\label{eq:scn2}
\end{eqnarray}
varying, again,
\begin{equation}
|\lambda|\,,|\kappa|\,; \ \
|A_\lambda|\,,|A_\kappa| \,; \ \
\phi_\kappa^\prime\,.
\end{equation}
We have fixed $M_1=M_2=-200$ GeV as in the previous case.
We observe that the effects of 
the CP phases $\phi_{A_t}$  and $\phi_{A_b}$ are negligible 
in this scenario and we simply take $\phi_{A_t}=\phi_{A_b}=0$.

\begin{figure}[t!]
\begin{center}
\includegraphics[width=5.2cm]{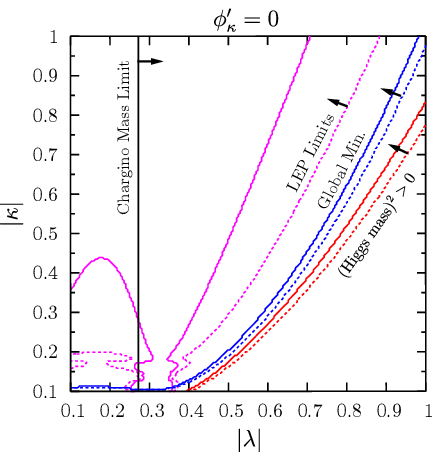}
%\hspace{0.5cm}
\includegraphics[width=5.2cm]{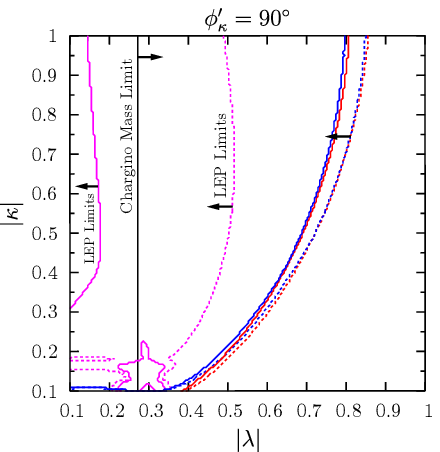}
%\\[0.5cm]
\includegraphics[width=5.2cm]{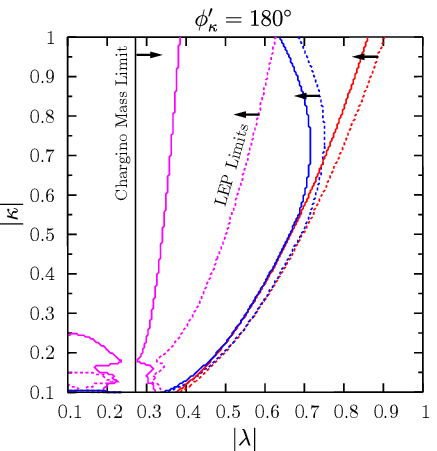}
\end{center}
\caption{The allowed region in the $|\kappa|$-$|\lambda|$ plane for the 
scenario in Eq.~(\ref{eq:scn2}) with
$|A_\lambda|=1200$ GeV and $|A_\kappa|=130$ GeV for
the three values of $\phi^\prime_\kappa=0^\circ\,,
90^\circ$, and $180^\circ$.
The lines are the same as in Fig.~\ref{fig:kap_lam_C1}.}
\label{fig:kap_lam_C2}
\end{figure}
\begin{figure}[t!]
\begin{center}
\includegraphics[width=5.3cm]{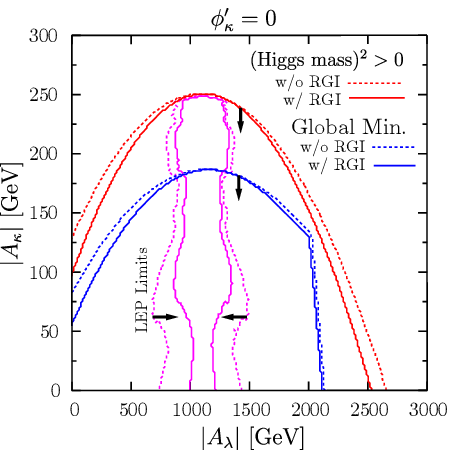}
%\hspace{0.5cm}
\includegraphics[width=5.3cm]{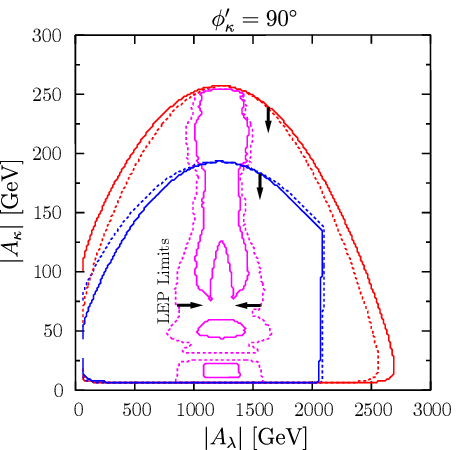}
%\\[0.5cm]
\includegraphics[width=5.3cm]{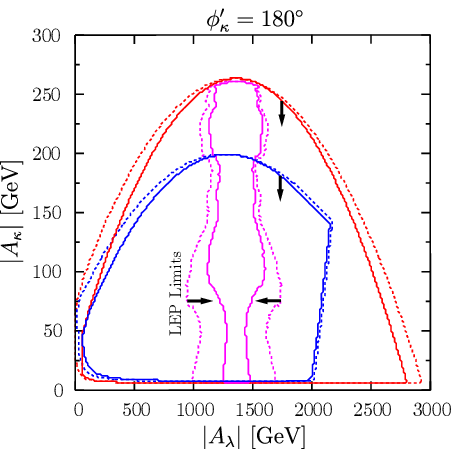}
\end{center}
\caption{The allowed region in the $|A_\kappa|$-$|A_\lambda|$ plane 
for the scenario~(\ref{eq:scn2}) 
with $|\lambda|=0.3$ and $|\kappa|=0.15$ for
the three values of $\phi^\prime_\kappa=0^\circ\,,
90^\circ$, and $180^\circ$.
The lines are the same as in Fig.~\ref{fig:kap_lam_C1}.}
\label{fig:Akap_Alam_C2}
\end{figure}
In Fig.~\ref{fig:kap_lam_C2}, we show the allowed region in 
the $|\kappa|$-$|\lambda|$ plane taking
$|A_\lambda|=1200$ GeV and $|A_\kappa|=130$ GeV for
the three values of $\phi^\prime_\kappa=0^\circ\,,
90^\circ$, and $180^\circ$.
We again see that
 the LEP limits, the global minimum, and the positive $M_H^2$ conditions
become stronger with the RG improvement. Among them, the LEP limits are most strongly
constraining the parameter space.
Especially, when $\phi^\prime_\kappa=90^\circ$ only a small region
with $0.25\lsim |\lambda| \lsim 0.35$ and $|\kappa|\lsim 0.2$ is allowed 
after including the chargino mass limit.
In Fig.~\ref{fig:Akap_Alam_C2}, we show the allowed region in 
the $|A_\kappa|$-$|A_\lambda|$ plane 
taking $|\lambda|=0.3$ GeV and $|\kappa|=0.15$.
The allowed range of $|A_\lambda|$ is around $\sim 1100$ GeV when
$\phi^\prime_\kappa=0^\circ$ and it moves to a higher-value region as
$\phi^\prime_\kappa$ increases, similar to 
the scenario considered in the previous section.
The LEP limits and the positivity condition of $M_H^2>0$ 
constrain the parameter space
$|A_\kappa| \lsim 250$ GeV, and it is further constrained to 
$|A_\kappa| \lsim 200$ GeV by the global minimum condition.
Additional restriction arises for small $|A_\kappa|$ when
$\phi'_\kappa = 90^\circ, 180^\circ$.

\begin{figure}[t!]
\begin{center}
\includegraphics[width=7.0cm]{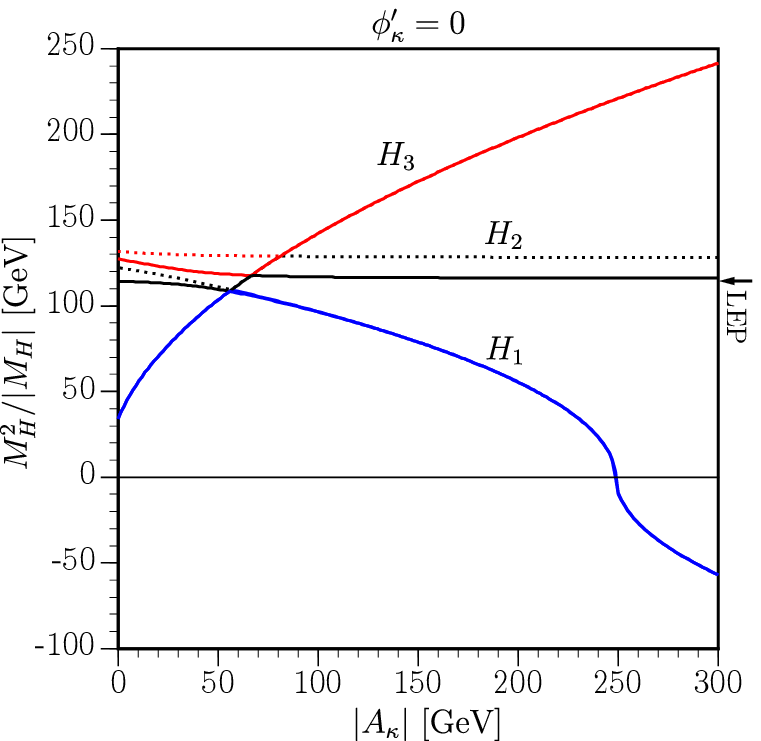}
\hspace{0.5cm}
\includegraphics[width=6.6cm]{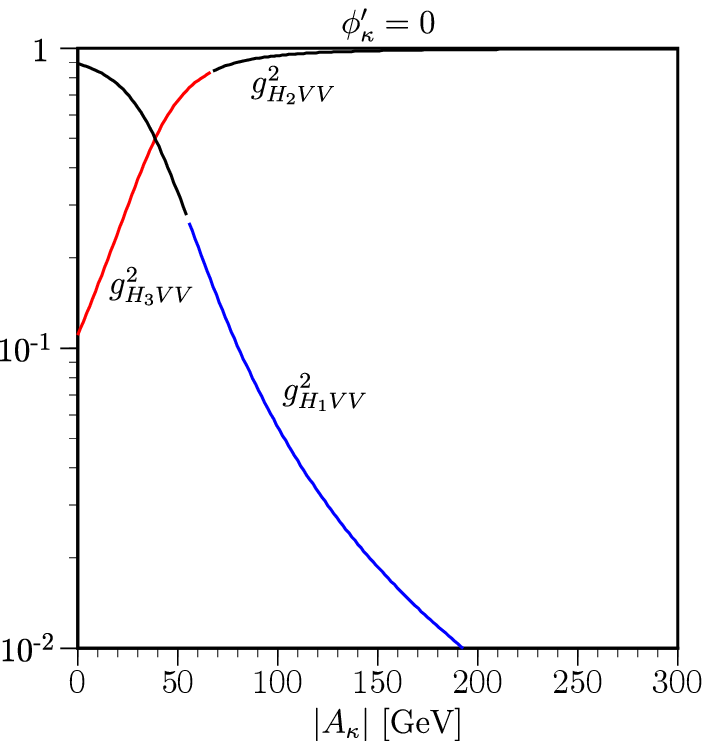}
\\[0.5cm]
\includegraphics[width=7.0cm]{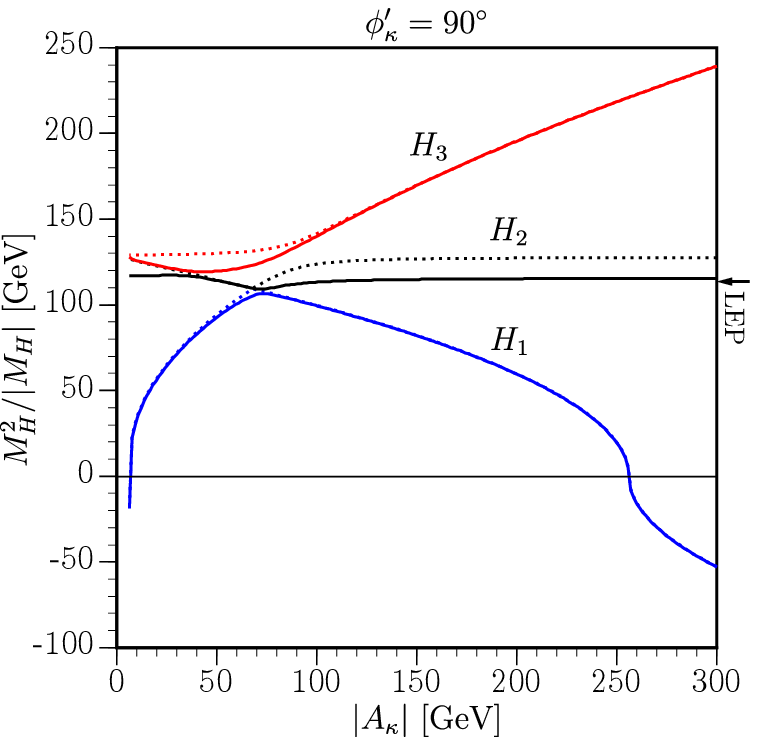}
\hspace{0.5cm}
\includegraphics[width=6.6cm]{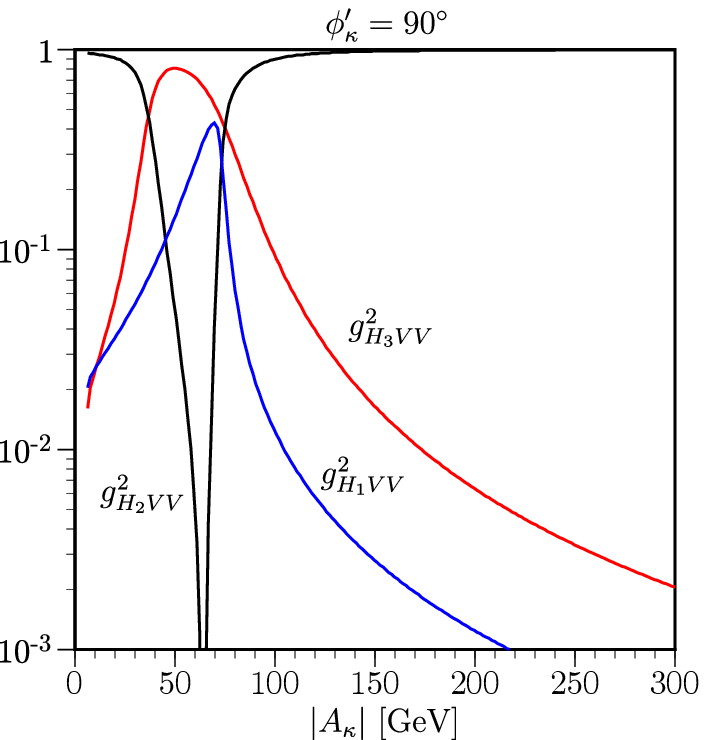}
\end{center}
\caption{ The masses $M_{H_i}^2/|M_{H_i}|$ (left panels) and 
couplings $g^2_{H_iVV}$ (right panels) for $i=1,2,3$ as functions of
$|A_\kappa|$ taking $|A_\lambda|=1200$ GeV
for the scenario in Eq.~(\ref{eq:scn2}).
The global minimum condition constrains $|A_\kappa|$ as
$0\leq|A_\kappa| \lsim 185$ GeV ($\phi^\prime_\kappa=0^\circ$, upper panels) and
7 GeV $ \lsim |A_\kappa| \lsim 192$ GeV ($\phi^\prime_\kappa=90^\circ$, lower panels).}
\label{fig:mhghvv1_C2}
\end{figure}
To understand why the LEP limits become stronger with
the RG improvement and why the allowed region shows interesting
features when $\phi^\prime_\kappa=90^\circ$ 
(see the middle frame of Fig.~\ref{fig:Akap_Alam_C2}),
we examine the three 
light Higgs-boson masses and their couplings to a pair of vector bosons.
In Fig.~\ref{fig:mhghvv1_C2}, we show the masses (left panel)
and couplings (right panel)  as functions of $|A_\kappa|$ taking
$|A_\lambda|=1200$ GeV. In the left panels, the dotted lines 
are for the masses without the RG improvement.
The lightest Higgs-boson mass becomes negative 
when $|A_\kappa| \gsim$ 250 GeV, signaling
the tachyonic state.
For $\phi^\prime_\kappa=0$ (upper panels), two level crossings happen
at $|A_\kappa|=50$ GeV and $\sim 70$ GeV. Below $|A_\kappa|\lsim 50$ GeV,
$H_1$ (blue lines) is CP odd and the RG improvement pushes down 
the mass of $H_2$ (black lines) to make the LEP limit stronger.
When $|A_\kappa|$ is between $\sim 50$ GeV and $\sim 70$ GeV, 
$H_2$ is CP odd and $H_1$ is near to the LEP limit with sizable coupling
$g^2_{H_1VV} \gsim 10^{-1}$. When $|A_\kappa|\gsim 70$ GeV, $H_3$ is CP odd
and the RG improvement decreases the mass of $H_2$ by the amount of $\sim 10$ GeV but 
$M_{H_2}$ is still above the LEP limit, though very near to it.
When $\phi^\prime_\kappa=90^\circ$ (lower panels), the three states do not carry 
definite CP parities and $g^2_{H_2VV}$ is the largest when
$|A_\kappa|$ is below 35 GeV and above 75 GeV, explaining the large correction to
$M_{H_2}$ (black) there.
Between $|A_\kappa|=35$ GeV and 75 GeV, $g^2_{H_3VV}$ is the largest and
$M_{H_3}$ (red) is affected by the RG improvement most significantly.
We note that $g^2_{H_1VV}$ is enhanced in the region
around $|A_\kappa|=70$ GeV. This explains why the region is ruled out 
by the LEP limits, while allowed in the CP-conserving case.

\begin{figure}[t!]
\begin{center}
\includegraphics[width=7.0cm]{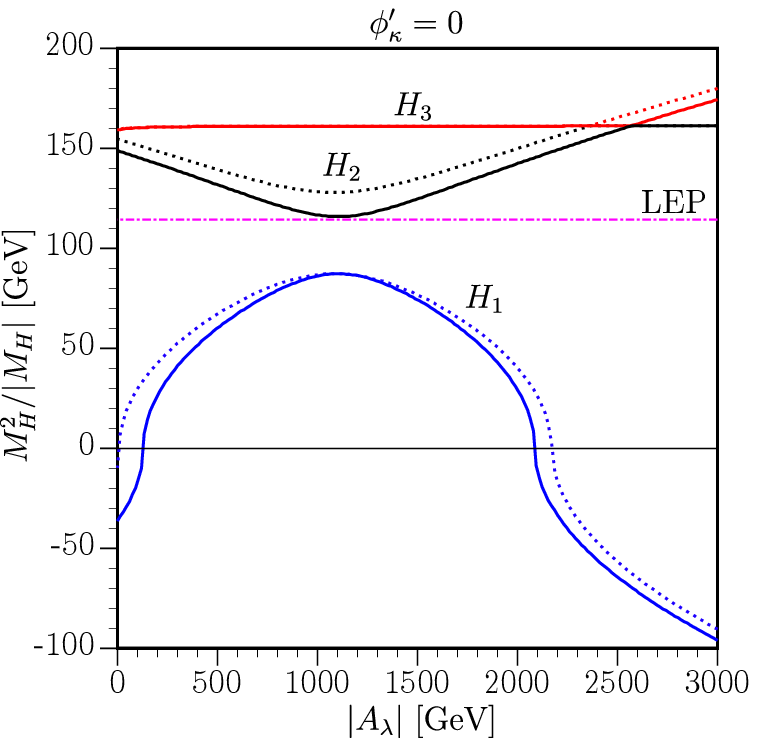}
\hspace{0.5cm}
\includegraphics[width=6.6cm]{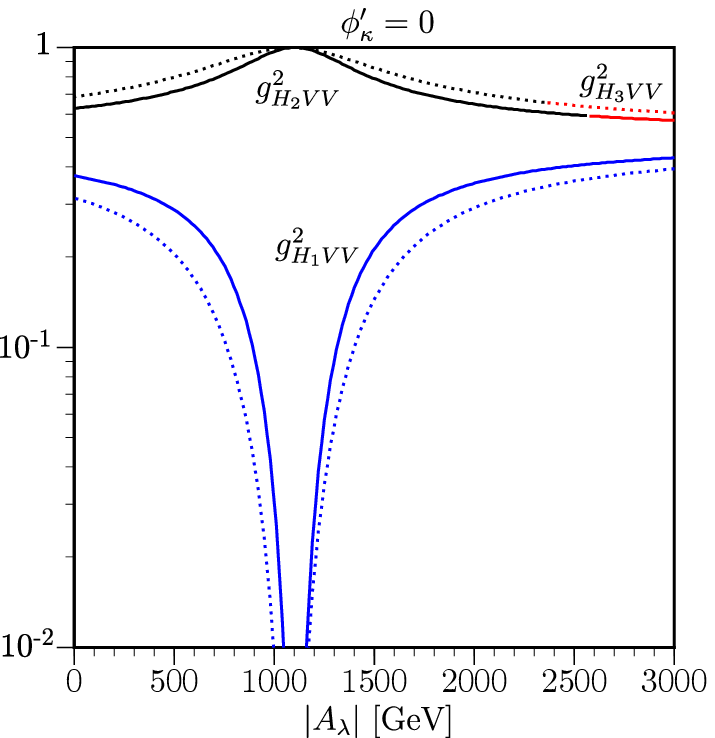}
\\[0.5cm]
\includegraphics[width=7.0cm]{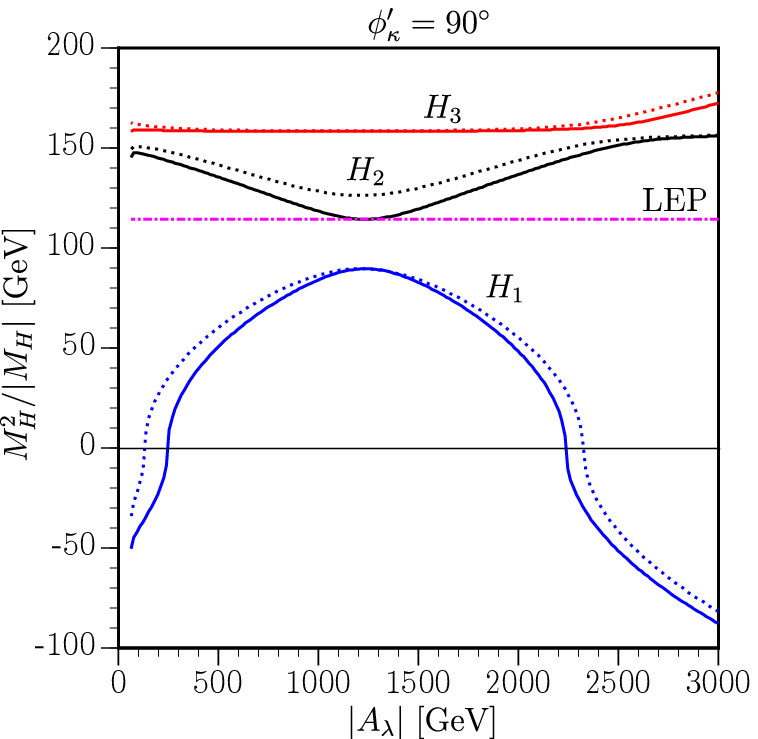}
\hspace{0.5cm}
\includegraphics[width=6.6cm]{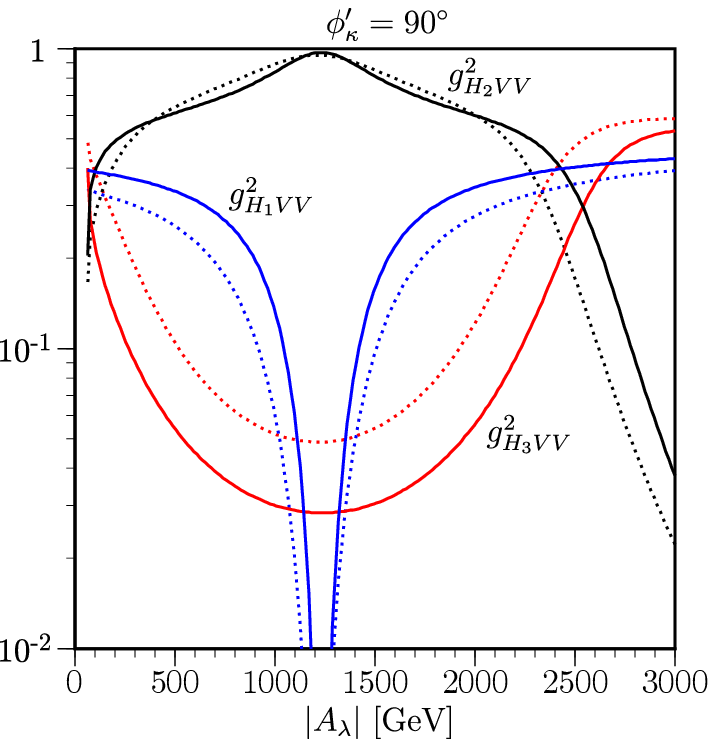}
\end{center}
\caption{The masses $M_{H_i}^2/|M_{H_i}|$ (left panels) and 
couplings $g^2_{H_iVV}$ (right panels) for $i=1,2,3$ as functions of
$|A_\lambda|$ taking $|A_\kappa|=130$ GeV
for the scenario in Eq.~(\ref{eq:scn2}).
The global minimum condition constrains $|A_\lambda|$ as
400 (305) GeV $\lsim  |A_\lambda| \lsim $ 1995 (2018) GeV 
($\phi^\prime_\kappa=0^\circ$, upper panels) and
426 (328) GeV $\lsim |A_\lambda| \lsim $ 2073 (2092) GeV 
($\phi^\prime_\kappa=90^\circ$, lower panels)
with (without) the RG improvement.
}
\label{fig:mhghvv2_C2}
\end{figure}
In Fig.~\ref{fig:mhghvv2_C2}, we show the masses (left panels)
and the couplings (right panels) of the light three Higgs bosons 
as functions of $|A_\lambda|$ taking $|A_\kappa|=130$ GeV.
Around $|A_\lambda| \sim 1200$ GeV,
in both cases with $\phi^\prime_\kappa=0^\circ$ and $90^\circ$,
$g_{H_2VV}^2$ (black lines) is the largest and
the RG improvement decreases the mass of $H_2$ and increases the mixing
between $H_1$ and $H_2$ states, making the LEP limits stronger.
We observe that the LEP limits allows only the region around
$|A_\lambda| \sim 1200$ GeV since when 
moving away from the point,
the mass of $H_1$ decreases while its coupling is rapidly increasing.

%-------------------
\subsection{An electroweak baryogenesis (EWBG)-motivated scenario}
%-------------------
%

The last scenario we are considering has an intermediate value of $\tan\beta$ with
small $v_S$:
\begin{eqnarray}
&&
\tan\beta=5\,, \ \
v_S=200~{\rm GeV}\,,
\nonumber \\
&&
M_{\widetilde Q}=M_{\widetilde U}=M_{\widetilde D}=|A_t|=|A_b|=1000~{\rm GeV}\,,
\nonumber \\
&&
\phi_\lambda^\prime=0\,, \ \
%\phi_{A_t}=\phi_{A_b}=0\,,
\nonumber \\
&&
{\rm sign}\,[\cos(\phi^\prime_\kappa+\phi_{A_\kappa})] =
{\rm sign}\,[\cos(\phi^\prime_\lambda+\phi_{A_\lambda})]  = + 1 \,,
\label{eq:scn3}
\end{eqnarray}
while varying 
\begin{equation}
|\lambda|\,,|\kappa|\,; \ \
|A_\lambda|\,,|A_\kappa| \,; \ \
\phi_\kappa^\prime\,,\phi_A\,,
\end{equation}
where $\phi_A\equiv \phi_{A_t}=\phi_{A_b}$ denotes the 
common CP phase of the third-generation trilinear terms.
If it is not mentioned otherwise, we are taking
$\phi_A=0$.
We have fixed $M_1=M_2=-200$ GeV as in the previous cases. 
We find that a first-order phase transition could occur in some
regions of the parameter space of this scenario~\cite{Funakubo:2005pu},
which is needed
for the EWBG~\cite{ewbg}.

\begin{figure}[t!]
\begin{center}
\includegraphics[width=5.3cm]{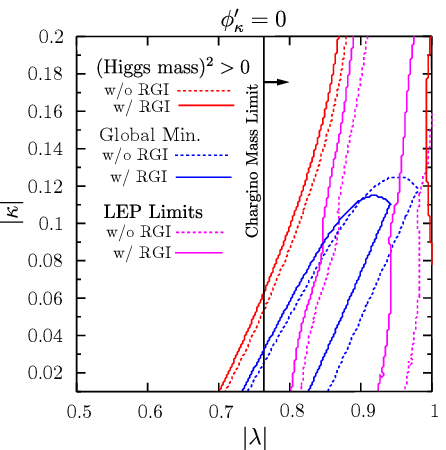}
%\hspace{0.5cm}
\includegraphics[width=5.3cm]{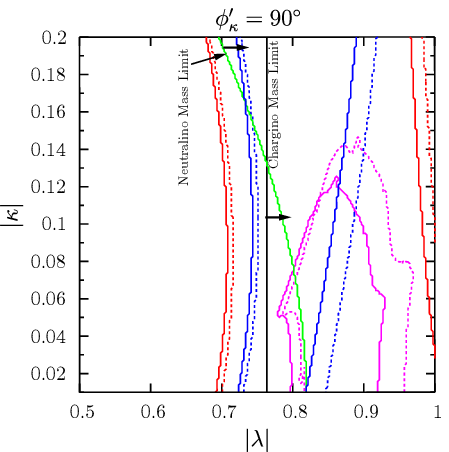}
%\hspace{0.5cm}
\includegraphics[width=5.3cm]{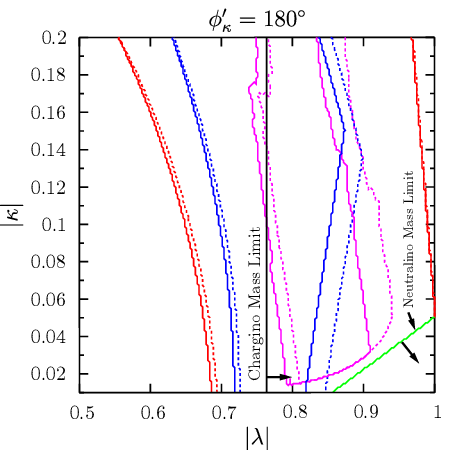}
\end{center}
\caption{The allowed region in the $|\kappa|$-$|\lambda|$ plane
for the scenario in Eq.~(\ref{eq:scn3}) with
$|A_\lambda|=600$ GeV and $|A_\kappa|=125$ GeV. 
The lines are the same as in Fig.~\ref{fig:kap_lam_C1}.
}
\label{fig:kap_lam_C3}
\end{figure}
\begin{figure}[t!]
\begin{center}
\includegraphics[width=5.3cm]{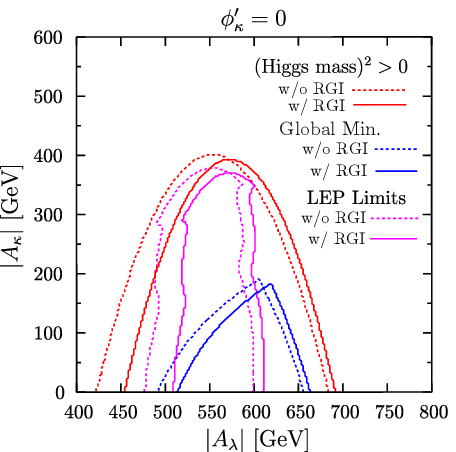}
%\hspace{0.5cm}
\includegraphics[width=5.3cm]{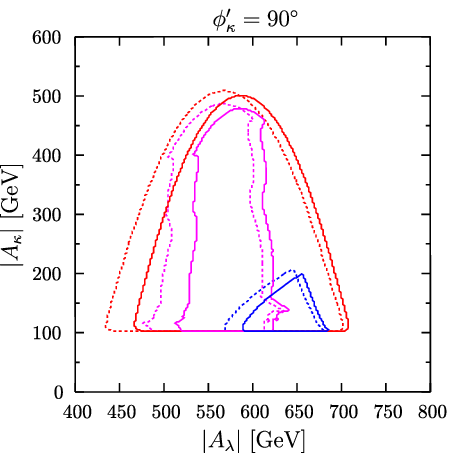}
%\hspace{0.5cm}
\includegraphics[width=5.3cm]{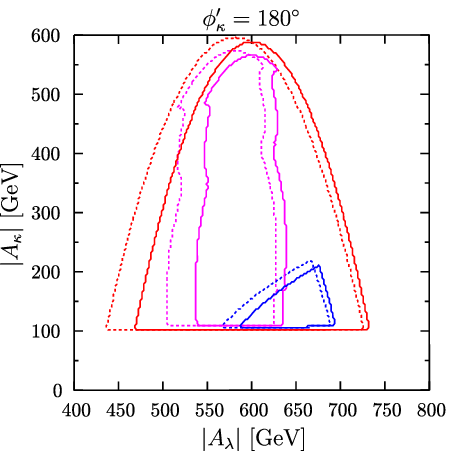}
\end{center}
\caption{The allowed region in the $|A_\kappa|$-$|A_\lambda|$ plane
for the scenario in Eq.~(\ref{eq:scn3}) with
$|\lambda|=0.83$ and $|\kappa|=0.05$.
The lines are the same as in Fig.~\ref{fig:kap_lam_C1}.
}
\label{fig:Akap_Alam_C3}
\end{figure}
In Fig.~\ref{fig:kap_lam_C3}, we show the allowed region in the 
$|\kappa|$-$|\lambda|$ plane taking $|A_\lambda|=600$ GeV and 
$|A_\kappa|=125$ GeV for the three values of
$\phi^\prime_\kappa=0^\circ$ (left panel), $90^\circ$ (middle panel), and $180^\circ$ (right panel).
Again, we observe that 
the allowed region is largely affected by the CP phase and
the RG improvement.  The RG improvement tends to
shift the allowed region to lower values of $|\lambda|$. 
The $|\kappa|$ is bounded above
by $\sim 0.12$ when $\phi^\prime_\kappa=90^\circ$.
In Fig.~\ref{fig:Akap_Alam_C3}, the allowed region is shown in the
$|A_\kappa|$-$|A_\lambda|$ plane taking $|\lambda|=0.83$
\footnote{
It was shown in Ref.~\cite{Miller:2003ay} that a $|\lambda|=0.7$ at the
weak scale is perfectly safe with perturbativity below the grand unified theory (GUT)
scale, while we have verified that a $|\lambda| =0.8$ 
at the weak scale still gives a value
below $4\pi$ at the GUT scale. Therefore, the value $|\lambda|=0.83$
chosen is not expected to have any serious violation of
perturbativity up to the GUT scale.} and
$|\kappa|=0.05$ for the three values of
$\phi^\prime_\kappa=0^\circ$ (left panel),
$90^\circ$ (middle panel), and
$180^\circ$ (right panel).
The RG improvement tends to
shift the allowed region to higher values of $|A_\lambda|$ and
the global minimum condition is stronger when
$\phi^\prime_\kappa=90^\circ$ and $180^\circ$,
leaving a small allowed region with 
$100~{\rm GeV}\lsim |A_\kappa|\lsim 200$ GeV around $|A_\lambda|\sim 600$ GeV.

\begin{figure}[t!]
\begin{center}
\includegraphics[width=7.0cm]{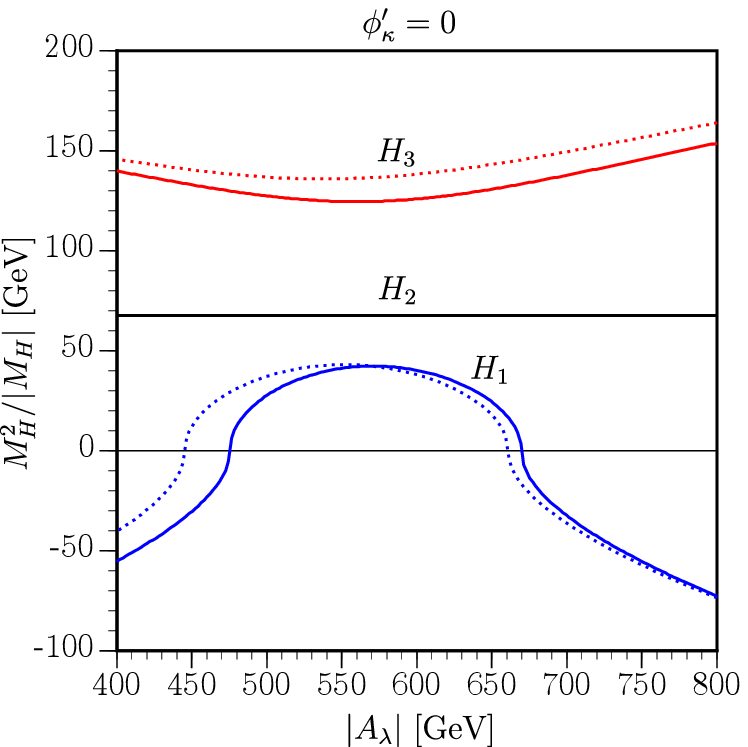}
\hspace{0.5cm}
\includegraphics[width=6.8cm]{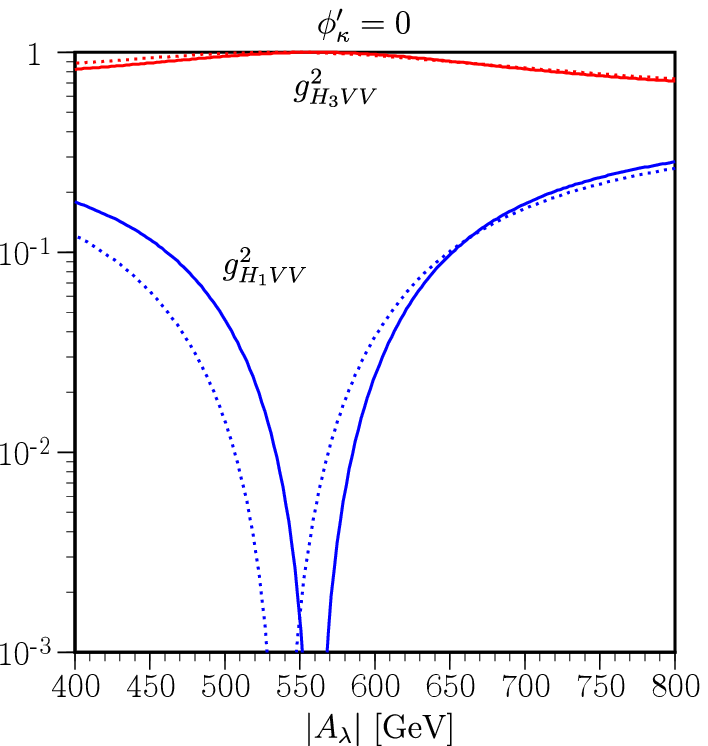}
\\[0.5cm]
\includegraphics[width=7.0cm]{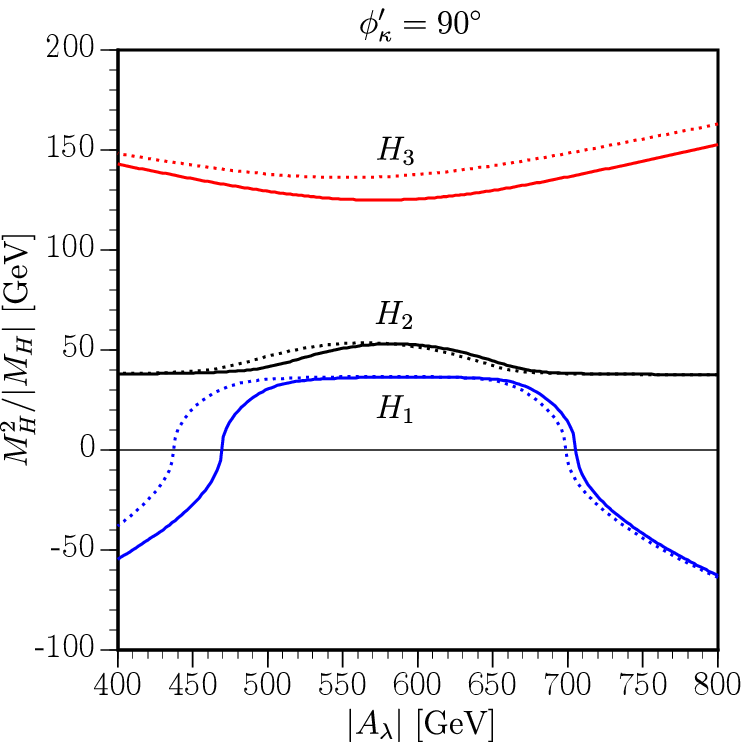}
\hspace{0.5cm}
\includegraphics[width=6.8cm]{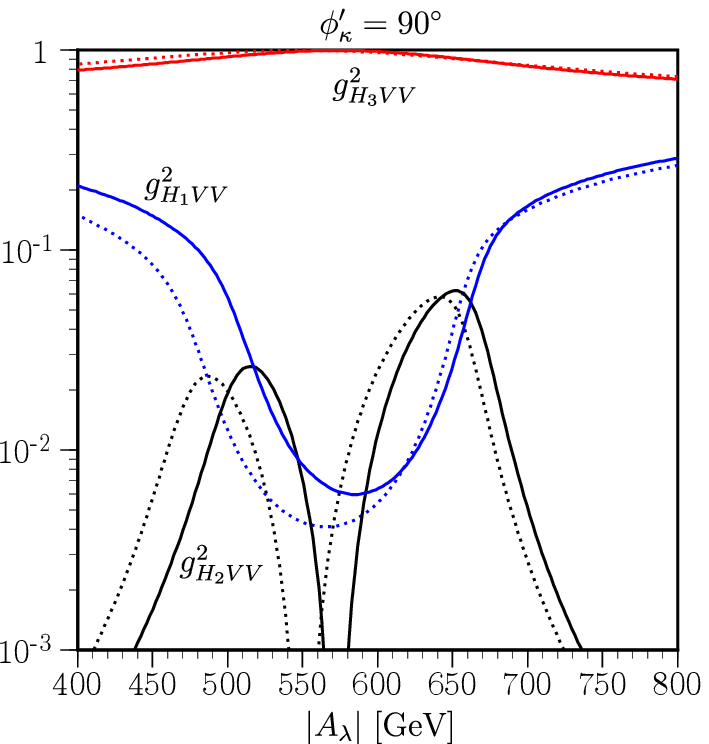}
\end{center}
\caption{ The masses $M_{H_i}^2/|M_{H_i}|$ (left panels) and 
couplings $g^2_{H_iVV}$ (right panels) as functions of
$|A_\lambda|$ taking $|A_\kappa|=125$ GeV
for the scenario in Eq.~(\ref{eq:scn3}) with
$|\lambda|=0.83$, $|\kappa|=0.05$.
The global minimum condition constrains $|A_\lambda|$ as
551 GeV $\lsim |A_\lambda| \lsim$ 625 GeV ($\phi^\prime_\kappa=0^\circ$ without the RGI),
571 GeV $\lsim |A_\lambda| \lsim$ 634 GeV ($\phi^\prime_\kappa=0^\circ$ with the RGI);
576 GeV $\lsim |A_\lambda| \lsim$ 668 GeV ($\phi^\prime_\kappa=90^\circ$ without the RGI),
596 GeV $\lsim |A_\lambda| \lsim$ 675 GeV ($\phi^\prime_\kappa=90^\circ$ with the RGI).
}
\label{fig:mhghvv1_C3}
\end{figure}
In Fig.~\ref{fig:mhghvv1_C3}, we show the masses (left panels)
and couplings (right panels)  as functions of $|A_\lambda|$ taking
$|A_\kappa|=125$ GeV. 
The dotted lines are for the masses and couplings 
without the RG improvement. We see that the RG improvement
shifts the masses and couplings to the region with
larger values of $|A_\lambda|$.

\begin{figure}[t!]
\begin{center}
\includegraphics[width=7.5cm]{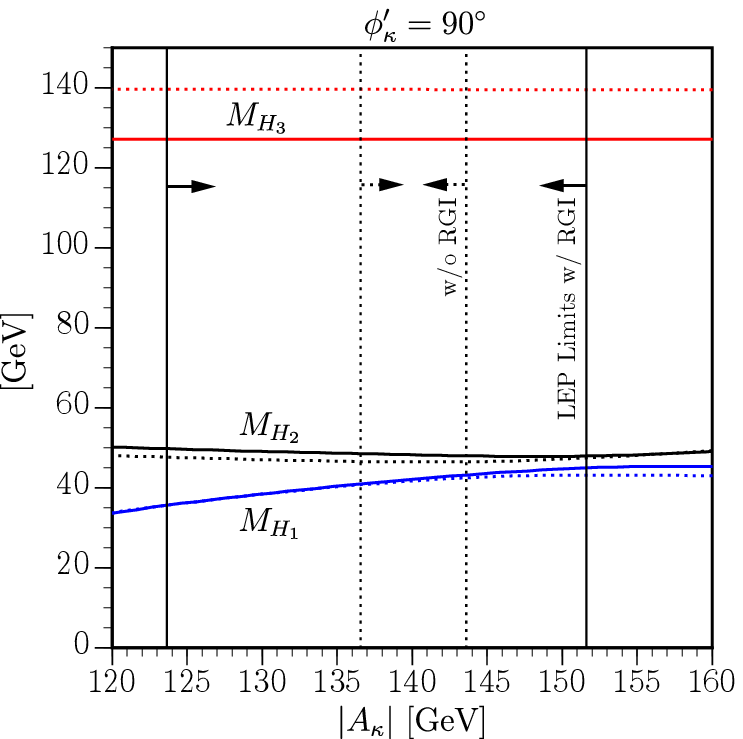}
\hspace{0.5cm}
\includegraphics[width=7.1cm]{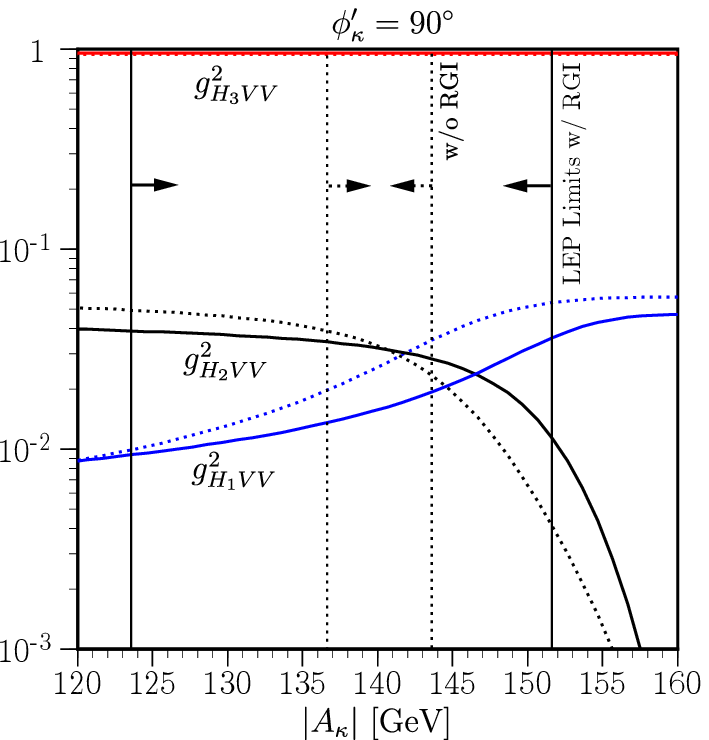}
\\[0.5cm]
\includegraphics[width=7.5cm]{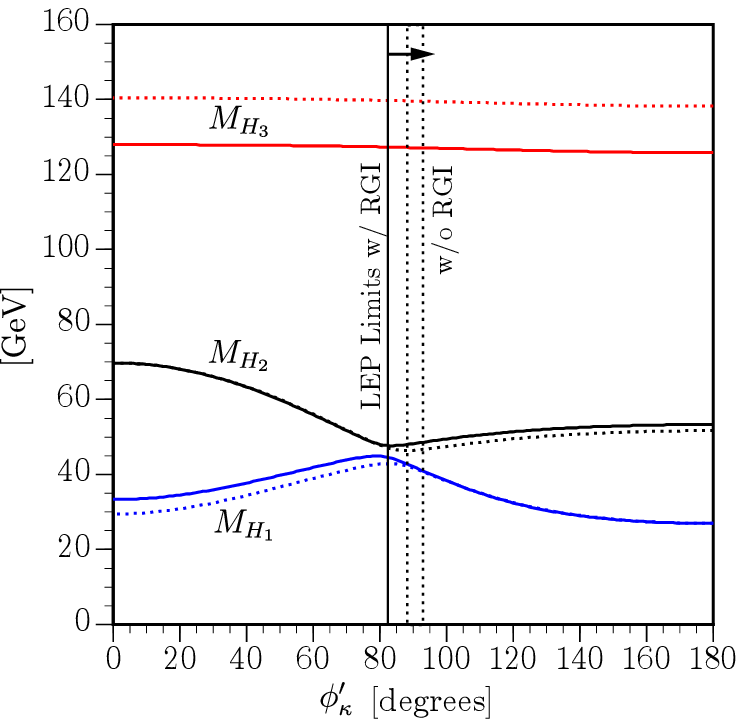}
\hspace{0.5cm}
\includegraphics[width=7.2cm]{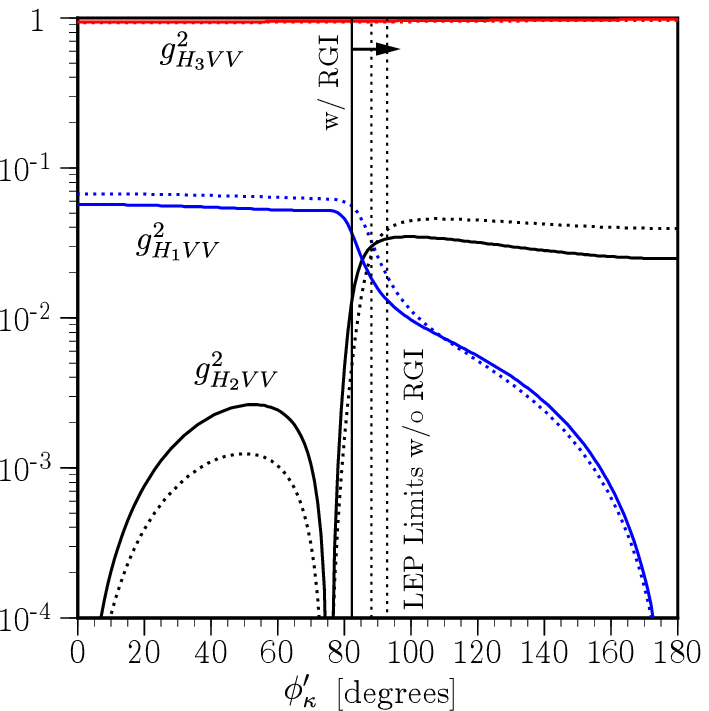}
\end{center}
\caption{ (Upper panels) The masses $M_{H_i}$ (left panels) and 
couplings $g^2_{H_iVV}$ (right panels) for $i=1,2,3$ as functions of
$|A_\kappa|$ taking $|A_\lambda|=625$ GeV and
$\phi^\prime_\kappa=90^\circ$
for the scenario in Eq.~(\ref{eq:scn3}) with
$|\lambda|=0.83$, $|\kappa|=0.05$.
(Lower panels) The same as in the upper panels but as functions of
$\phi^\prime_\kappa$ taking $|A_\kappa|=140$ GeV.
The vertical solid and dotted lines bound the 
allowed region with and without the RGI, respectively.
}
\label{fig:mhghvv2_C3}
\end{figure}
In Fig.~\ref{fig:mhghvv2_C3}, we show the masses (left panels)
and couplings (right panels)  as functions of $|A_\kappa|$ ($\phi^\prime_\kappa$)
at $\phi^\prime_\kappa=90^\circ$ ($|A_\kappa|=140$ GeV) in
the upper (lower) frames.
We have fixed $|A_\lambda|=625$ GeV, because for this value
the RG improvement could significantly enlarge the allowed region 
as seen from the middle and right panels of 
Fig.~\ref{fig:Akap_Alam_C3}.
Before including the RG improvement, as bounded by the
dotted vertical lines, the allowed region is very narrow:
\begin{eqnarray}
\begin{array}{clll}
137 {\rm ~GeV} \lsim |A_\kappa| \lsim 144{\rm ~GeV}  & {\rm ~when~} 
& \phi^\prime_\kappa=90^\circ &  {\rm~(upper)} \\[0.2cm]
88^\circ \lsim \phi^\prime_\kappa \lsim 93^\circ  & {\rm ~when~} 
& |A_\kappa|=140 {\rm ~GeV}  & {\rm ~(lower)}.
\end{array} \nonumber
\end{eqnarray}
While, including the RG improvement, the allowed region is enlarged as:
\begin{eqnarray}
\begin{array}{clll}
124 {\rm ~GeV} \lsim |A_\kappa| \lsim 152{\rm ~GeV}  & {\rm ~when~} 
& \phi^\prime_\kappa=90^\circ &  {\rm~(upper)} \\[0.2cm]
82^\circ \lsim \phi^\prime_\kappa \leq 180^\circ  & {\rm ~when~} 
& |A_\kappa|=140 {\rm ~GeV}  & {\rm ~(lower)}.
\end{array} \nonumber
\end{eqnarray}
This is because the couplings to the weak gauge bosons are 
on the verge of the LEP-allowed region and the size of the couplings
are generically reduced by the RG improvement,
as seen from the right panels of Fig.~12.

\begin{figure}[t!]
\begin{center}
\includegraphics[width=10.0cm]{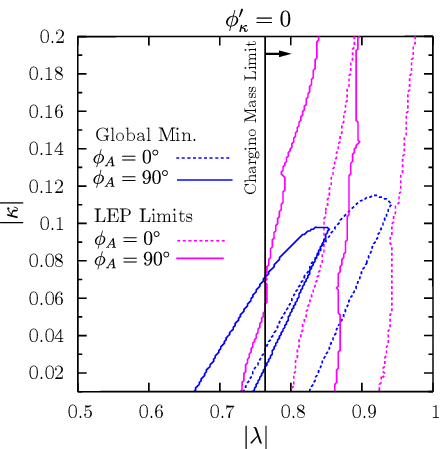}
%\hspace{0.5cm}
\includegraphics[width=8.0cm]{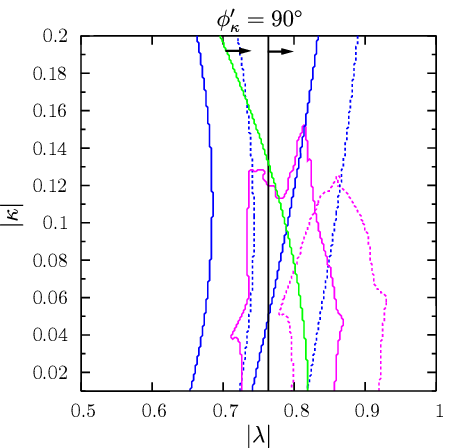}
%\\[0.5cm]
\includegraphics[width=8.0cm]{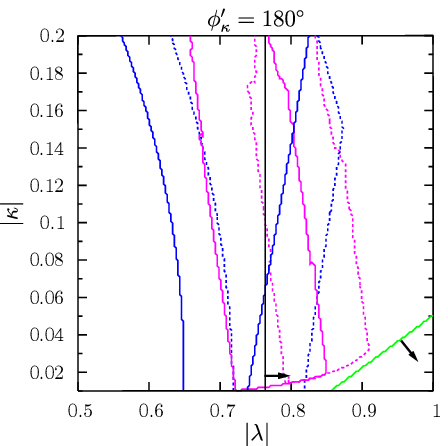}
\end{center}
\caption{The allowed region in the $|\kappa|$-$|\lambda|$ plane for 
the two values of $\phi_A=0^\circ$ (dashed lines) and $90^\circ$ (solid lines)
taking the scenario in Eq.~(\ref{eq:scn3}) with
$|A_\lambda|=600$ GeV, $|A_\kappa|=125$ GeV and
the three values of $\phi^\prime_\kappa=0^\circ$ (upper panel) and
$90^\circ$ and $180^\circ$ (lower panels). The RG improvement has been
included in all cases.
The dashed lines for $\phi_A=0^\circ$ 
are the same as the solid lines in Fig.~\ref{fig:kap_lam_C3}.}
\label{fig:kap_lam_C3_PHIA}
\end{figure}
\begin{figure}[t!]
\begin{center}
\includegraphics[width=7.5cm]{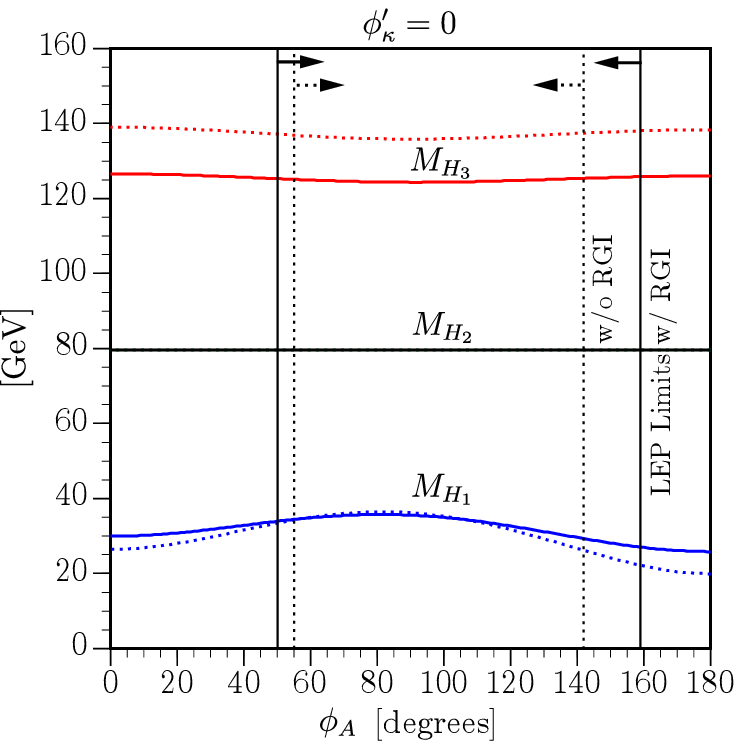}
\hspace{0.5cm}
\includegraphics[width=7.1cm]{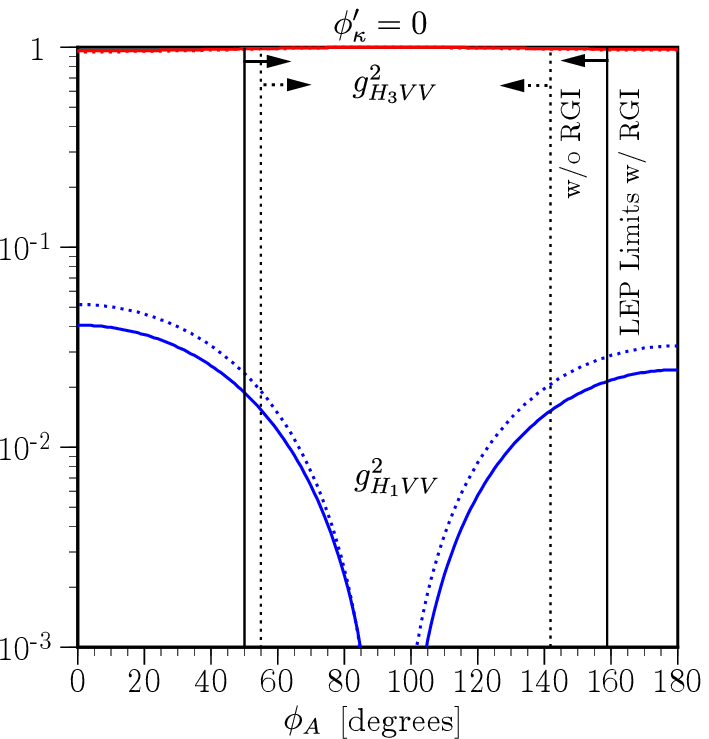}
\end{center}
\caption{ The masses $M_{H_i}$ (left panel) and
couplings $g^2_{H_iVV}$ (right panel) for $i=1,2,3$ as functions of $\phi_A$
taking $|A_\kappa|=125$ GeV and $|A_\lambda|=600$ GeV 
for the scenario in Eq.~(\ref{eq:scn3}) with
$|\lambda|=0.815$, $|\kappa|=0.07$, and 
$\phi^\prime_\kappa=0^\circ$.
The vertical solid and dotted lines bound the
allowed region with and without the RGI, respectively.
}
\label{fig:mhghvv3_C3}
\end{figure}
Finally, we study the effect of $\phi_A$, the common CP phase 
of the third-generation trilinear terms. Being different from the
previous LEP-compatible scenario in Eq.~(\ref{eq:scn2}), 
we find that the EWBG-motivated scenario in Eq.~(\ref{eq:scn3})
is sensitive to $\phi_A$.
In Fig.~\ref{fig:kap_lam_C3_PHIA}, we show the dependence 
of the allowed  region in the $|\kappa|$-$|\lambda|$ plane on $\phi_A$ 
taking $|A_\lambda|=600$ GeV and $|A_\kappa|=125$ GeV for
the three values of $\phi^\prime_\kappa=0^\circ$ (upper panel) and
$90^\circ$ and $180^\circ$ (lower panels).
The RG improvement has been included in all cases.
The solid  lines are for $\phi_A=90^\circ$ while
the dashed lines are for $\phi_A=0^\circ$ which
are the same as the solid lines in Fig.~\ref{fig:kap_lam_C3} with the RGI.
The bounding lines from the condition $M_H^2>0$ are not shown here
because the condition is always weaker than the LEP limits in this case.
We observe that the allowed regions prefer lower $|\lambda|$
values when $\phi_A=90^\circ$.
For $\phi^\prime_\kappa=0^\circ$ and $90^\circ$ there is no region in which
the global minimum condition and the LEP limits
can be satisfied simultaneously for both values of $\phi_A$.
In Fig.~\ref{fig:mhghvv3_C3}, we show the dependence 
of the Higgs masses and couplings on $\phi_A$ at
$\phi^\prime_\kappa=0^\circ$. This figure
illustrates the interesting case in which the CP-conserving
limits with $\phi_A=0^\circ$ and $180^\circ$ are not 
compatible with the LEP limits.
The specific parameter set chosen satisfies the LEP constraints
only with nontrivial $\phi_A$.
This is because a parametric cancellation occurs between
the two terms contributing to the coupling 
$g_{H_1VV}=O_{11}c_\beta+O_{21}s_\beta$
around $\phi_A=90^\circ$,
leading to the suppression of the coupling, as shown
in the right panel.

%%%%%%%%%%%%%%%%%%%%%-------------------
\section{Conclusions}
%%%%%%%%%%%%%%%%%%%%%-------------------
%
We have performed a comprehensive study on the mass spectrum, mixing,
and couplings to weak gauge bosons of the Higgs sector within the 
NMSSM, which is a gauge singlet extension of the MSSM
with $Z_3$ symmetry to address the $\mu$ problem.
The CP-violating parameters in the superpotential 
and in the soft SUSY-breaking terms
are fully taken into account.

At the tree level, there are three rephasing invariant 
combinations of the CP phases, two of which are fixed by the CP-odd
tadpole conditions up to a twofold ambiguity.
In contrast to the MSSM, there still remains one
physical CP phase, which induces the 
tree-level CP-violating mixing among the
5 neutral Higgs states.
Apart from the CP phase $(\phi^\prime_\lambda-\phi^\prime_\kappa)$,
the tree-level Higgs sector
is completely determined by the additional 7 real parameters: 
$(i)$ magnitudes of the two couplings, $|\lambda|$ and $|\kappa|$,
$(ii)$ the  three VEVs, $v_u$, $v_d$, and $v_S$
(or $v$, $\tan\beta$, and $v_S$), and
$(iii)$ magnitudes of the two $A$ terms, $|A_\lambda|$ and $|A_\kappa|$.
With the general notion of the tree-level CP-violating mixing in the
neutral Higgs sector, we derive a perturbative way to 
block-diagonalize a symmetric $(n+m)\times (n+m)$ matrix iteratively
and present analytic expressions for the leading-order 
effects for the CP-violating mixing when
the perturbative expansion of the mass matrix works reasonably.

We have computed the masses and mixing matrix of the Higgs bosons 
at one-loop level using the effective potential method.
We have taken into account the CP phases of the stop and sbottom 
sectors, which enter through the combinations 
of $\phi^\prime_\lambda+\phi_{A_t}$ and 
$\phi^\prime_\lambda+\phi_{A_b}$.
We also include the logarithmically enhanced 
two-loop corrections of the order ${\cal O}(g_s^2 h^4)$
and ${\cal O}(h^6)$ by performing the RG improvement
of the one-loop effective potential,
which has been implemented only in the CP-conserving limit before.
Beyond tree level our results are in agreement with those
in the literature, in the CP-conserving limit \cite{nmhdecay}
and in the case of without renormalization-group improvement 
\cite{Funakubo:2004ka}.

In our numerical analyses, we have considered the following
three different scenarios:
\begin{eqnarray}
\begin{array}{lll}
{\rm S1}\;\mbox{(Typical)}: & \tan\beta=3\,, &  v_S=750 {\rm ~GeV} \\
{\rm S2}\;\mbox{(LEP-compatible)}: & \tan\beta=10\,, &  v_S=600 {\rm ~GeV} \\
{\rm S3}\;\mbox{(EWBG-motivated)}: & \tan\beta=5\,, &  v_S=200 {\rm ~GeV}
\end{array} \nonumber
\end{eqnarray}
We have chosen the phase convention with $\phi^\prime_\lambda=0$ and varied
$
|\lambda|\,,|\kappa|\,; \ \
|A_\lambda|\,,|A_\kappa| \,; \ \
\phi_\kappa^\prime\,,\phi_A=\Phi_{A_t,A_b}\,.
$
For the SUSY-breaking parameters we have fixed
$M_{\widetilde Q}=M_{\widetilde U}=M_{\widetilde D}=|A_t|=|A_b|=1~{\rm TeV}$ and,
for other parameters, we refer to Eqs.~(\ref{eq:scn1}), (\ref{eq:scn2}),
and (\ref{eq:scn3}).
In each scenario, the following three main conditions are imposed
to derive constraints on the parameter space:
$(i)$ the LEP limits,
$(ii)$ the global minimum condition, and
$(iii)$ the positivity of the Higgs-boson masses squared.
The third condition is always weaker than the other two.
The global minimum condition does not allow too large values for
the trilinear parameters $|A_\lambda|$ and $|A_\kappa|$ because
the energy of the presumed EW vacuum is proportional to them.
The LEP limits constrain the allowed parameter space around
$|A_\lambda| \sim |\lambda| v_S t_\beta /\sqrt{2}$, which is also
the typical size of the heavier Higgs bosons.
The relative strength of the global minimum condition
and the LEP limits depends on the scenarios.

The renormalization-group improvement included in this study 
substantially strengthens the LEP limits, thus making it
more restrictive.
In the typical scenario {\rm S1}, the allowed region of the parameter
space strongly depends on the CP phase $\phi^\prime_\kappa$ and the RG
improvement. We found that the typical points with small
$|\lambda|$ and $|\kappa|$, which are allowed before the
inclusion of the RG improvement, are completely ruled out 
by the LEP limits after including the RG improvement.
In the LEP-compatible scenario {\rm S2},
the allowed region of the parameter space also strongly 
depends on the CP phase $\phi^\prime_\kappa$ and the RG
improvement  but the dependence on
$\phi_A$ is weak. We observe that the RG-improved
correction reduces the mass of the SM-like Higgs boson by an
amount of a few GeV to about 10 GeV, increases the mixing between the lighter states,
and shrinks the allowed parameter space significantly. 
When $\phi^\prime_\kappa$ takes on nontrivial values,
the lighter states do not carry definite CP parities and the
shape of the allowed parameter region becomes more complicated compared to the
CP-conserving case.
Last, in the EWBG-motivated scenario {\rm S3}, we find that the
global minimum condition restricts the parameter space more tightly
than the LEP limits and some parameter region, which is not allowed
in the CP-conserving case, could be allowed by assuming nontrivial
values of $\phi^\prime_\kappa$ and $\phi_A$, enlarging the allowed
parameter space.

We offer a few more comments before closing such as the following:
\begin{enumerate}
\item This is the first time that the next-to-minimal supersymmetric
standard model is studied allowing CP phases in the $\mu$ and soft 
SUSY-breaking parameters, and including full one-loop corrections with
renormalization-group improvement.  Substantial corrections to the Higgs-boson spectrum, 
mixing, and couplings to weak gauge bosons are realized,
Furthermore, nontrivial variations in the mass spectrum, mixing,
and couplings appear due to nonzero CP phases.
Therefore, we anticipate a whole new set of phenomenology associated
within this CP-violating NMSSM framework. 

\item It is well known that the experimental measured EDMs place
nontrivial constraint on the CP phases.  With one more physical 
CP phase added in this NMSSM framework the predictions for EDMs
are important to constrain the combinations of phases \cite{future1}.
Such CP phases are also important to provide enough CP violation
required in the electroweak baryogenesis.

\item A successful supersymmetry model should be able to explain
the anomalous magnetic moment of the muon, which is widely accepted
as a $3\sigma$ effect \cite{hagiwara}.  The CP-violating NMSSM considered
in this work should also be constrained so as to satisfy the muon
anomaly.  It is a nontrivial extension in this regard because of
the presence of many new Higgs bosons, which can be very light and
with CP violating couplings. 

\item There are a number of low-energy constraints on the lightest CP-odd
Higgs boson in the CP-conserving NMSSM \cite{hiller}. The exercises
can be repeated in the presence of the new CP phases.

\item There are a number of important cubic terms in the Higgs potential,
which also have nontrivial dependence on the Higgs spectrum and CP phases.
Specifically, for successful baryogenesis,
the soft cubic term involving the singlet field
$\lambda A_\lambda SH_dH_u + {\rm h.c.}$, which is absent in the MSSM,
is vital to enable a first-order phase transition
when the stops are heavy. 
Apparently, from the EWBG-motivated scenario studied in this work,
a first-order
phase transition is possible in this framework.  We will delay
this issue to a detailed study in the future \cite{future1}

\item A whole new set of phenomenology has to be explored in 
the Higgs sector with 5 neutral Higgs bosons with no definite
CP parities, and a pair of charged Higgs bosons.  As we can
see in this work, the couplings to weak gauge bosons vary
nontrivially with the CP phases.  The same can be said for 
the mass spectrum. It is more complicated
than the CP-violating MSSM or CP-conserving NMSSM.
At this point, we cannot forecast how the
decay branching ratios and production will be modified.
We delay this to a further study.

\item
With CP violation there is no explicit CP property for the Higgs
bosons.  In most of the cases, there are three relatively light
neutral Higgs bosons and two relatively heavy ones.  
The collider phenomenology is particularly concerned with the 
three lighter ones, 
in which one of them is the SM-like Higgs boson with a relatively large
coupling to the gauge bosons than the other two Higgs bosons (but
still the strength is a fraction of the SM value.)  At the Tevatron,
the most useful Higgs production channel is via the associated production with a
$W$ or a $Z$ boson.  The production rate of the
SM-like Higgs boson is smaller than the corresponding SM Higgs boson, 
but it may be possible to produce more
than one Higgs boson.  At the LHC, on the other hand,
production is dominated by the gluon fusion.
The crucial strategies for Higgs-boson search depend on the decay pattern of the Higgs bosons.  

\item 
There are a number of possible channels that the SM-like Higgs
boson can decay into, including the dominant $b\bar b$, $\tau^+ \tau^-$,
the rare ones $\mu^+\mu^-$, $\gamma\gamma, Z\gamma$, and the possible
new ones $h_1 h_1$, $h_1 h_2$, and $h_2 h_2$ depending on the mass
spectrum. In the CP-conserving NMSSM, the SM-like Higgs boson can
decay into 2 CP-even lighter Higgs bosons or 2 CP-odd Higgs bosons,
but not a mixture.  Now with CP violation, the SM-like Higgs boson can
decay into $h_1  h_2$, which is not possible in the CP-conserving
case.  It is a clean signal of CP violation.  Further decays of $h_1$
and $h_2$ will give a total of four fermions in the final state, e.g.,
$4b$, $2b 2\tau$, $4\tau$, $4\mu$, $2\mu2\tau$, $2b2\mu$, etc. 
Feasibility and coverage of parameter space certainly deserve further studies.
\end{enumerate}

We conclude by summarizing that we have started a new avenue
in the CP-violating NMSSM, which involves a whole new set
of phenomenology in low-energy precision measurements, in 
the LHC Higgs-boson searches, baryogenesis, etc., to be
explored in the future.

\vspace{-0.2cm}
\subsection*{Acknowledgements} 
\vspace{-0.3cm}
\noindent
We thank Koichi Funakubo for helpful discussions.  
The work was  supported in parts
by the NSC of Taiwan (96-2628-M-007-002-MY3), the NCTS, 
and by the WCU program through the KOSEF funded by
the MEST (R31-2008-000-10057-0).

%
%\newpage
%\section*{Appendices}

\def\theequation{\Alph{section}.\arabic{equation}}
\begin{appendix}
%------------------------------------------------------
\setcounter{equation}{0}
\section{Appendix}
\label{sec:appendix}
%------------------------------------------------------
In this Appendix, we consider a block diagonalization of a
symmetric $(n+m)\times (n+m)$ matrix
\begin{equation}
S = \left(\begin{array}{cc}
A & C \\
C^T & B
\end{array} \right)\,,
\end{equation}
where the submatrices can be expanded as
\begin{equation}
A \equiv A_0 +\sum_{n=1} \epsilon^n A_n \ \ ; \ \
B \equiv \sum_{n=1} \epsilon^n B_n \ \ ; \ \
C \equiv \sum_{n=1} \epsilon^n C_n \,.
\end{equation}
Note that, for successful diagonalization, we require
the off-diagonal $n\times m$
submatrix $C$ and the lower
$m\times m$ submatrix $B$ to be suppressed by, at least,
one power of $\epsilon$ compared with the upper
diagonal $n\times n$ submatrix $A$.
The block diagonalization can be implemented
by introducing the mixing matrix
\begin{equation}
V = \left(\begin{array}{cc}
{\bf 1}_{n\times n}+y & x \\
-x^T & {\bf 1}_{m\times m}+z
\end{array} \right)\,,
\end{equation}
where the submatrices in the diagonal parts
are symmetric, $y^T=y$ and $z^T=z$, and all three submatrices can
also be expanded as
\footnote{Note that $y$ and $z$ starts from 
the second order of $\epsilon$, required by 
the $\epsilon$ expansion of the off-diagonal part $x$ of $V$
and the orthogonality of the mixing matrix $V$, as shown below.}
\begin{equation}
x \equiv \sum_{n=1} \epsilon^n x_n \ \ ; \ \
y \equiv \sum_{n=2} \epsilon^n y_n \ \ ; \ \
z \equiv \sum_{n=2} \epsilon^n z_n \,.
\end{equation}
Using the orthogonality of the matrix $V$, 
$VV^T=V^TV={\bf 1}_{(n+m)\times (n+m)}$, we impose the relations
\begin{eqnarray}
2y+y^2+xx^T = {\bf 0}_{n\times n} \ \ ; \ \
2z+z^2+x^Tx = {\bf 0}_{m\times m}  \ \ ; \ \
yx = xz \,,
\end{eqnarray}
which can be used to determine the diagonal matrices
$y$ and $z$ in terms of $x$ iteratively, 
order by order in $\epsilon$, as follows:
\begin{eqnarray}
y_2 &=& -\frac{1}{2}\,x_1 x_1^T \nonumber \\
y_3 &=& -\frac{1}{2}\,(x_1 x_2^T +x_2 x_1^T) \nonumber \\
y_4 &=& -\frac{1}{2}\,(x_1 x_3^T +x_3 x_1^T+x_2 x_2^T) 
-\frac{1}{8} (x_1 x_1^T)^2 \nonumber \\
&\cdots &  \nonumber \\
y_i &=& y_i(x_1, x_2, \cdots , x_{i-1})
\end{eqnarray}
and 
\begin{eqnarray}
z_{i} &=& y_i\left(x_j \to x_j^T ~{\rm for}~ j<i\right)
\end{eqnarray}
Therefore, 
to determine the diagonal entries of
the mixing matrix $V$ up to order $\epsilon^i$,
all we need to know is $(x_1, x_2, \dots , x_{i-1})$
which can be obtained by
requiring the off-diagonal part of $VSV^T$ to vanish,
order by order in $\epsilon$ up to $\epsilon^{i-1}$, and
in terms of the submatrices $A_{0,1,...,(i-1)}$, 
$B_{1,...,(i-1)}$, and $C_{1,2,...,(i-1)}$.

More specifically we define the block-diagonal matrix
\begin{equation}
\tilde{S} \equiv VSV^T=
\left(\begin{array}{cc}
\tilde{S}_{11} & \tilde{S}_{12} \\
\tilde{S}_{12}^T & \tilde{S}_{22}
\end{array} \right)\,,
\end{equation}
with
\begin{eqnarray}
\tilde{S}_{11} &=& A + (Ay+yA+Cx^T+xC^T) + xBx^T
+ (yAy+yCx^T+xC^Ty)
\nonumber \\
\tilde{S}_{22} &=& B + (x^TAx-C^Tx-x^TC) + (Bz+zB) +
(-zC^Tx-x^TCz) + zBz
\end{eqnarray}
in order of increasing power in $\epsilon$
of the leading  terms.
The vanishing off-diagonal part is
\begin{eqnarray}
\tilde{S}_{12} &=& (-Ax+C) + xB+ (-yAx-xC^Tx+yC+Cz) + 
xBz+yCz \;,
\end{eqnarray}
where the first, the second, the third, the fourth,
and the fifth term starts from
$\epsilon^1$, $\epsilon^2$, 
$\epsilon^3$, $\epsilon^4$, and $\epsilon^5$, respectively.
To solve $\tilde{S}_{12}={\bf 0}_{n\times m}$, we have made the following
rearrangement:
\begin{equation}
\tilde{S}_{12} = \sum_{i=1} \epsilon^i \left(\tilde{S}_{12}\right)_i
\equiv \sum_{i=1} \epsilon^i \left(-A_0 x_i+C_i +D_i\right) \;,
\end{equation}
where $D_i$'s are  functions of $A$, $B$, $C$, and
$(x_1, x_2, \dots , x_{i-1})$:
\begin{equation}
D_i=D_i(x_j;A_j,B_j,C_j)~{\rm with}~ j<i \;.
\end{equation}
Therefore, $(\tilde{S}_{12})_i={\bf 0}_{n\times m}$ can be solved 
iteratively to give $x_i$:
\begin{equation}
x_i=A_0^{-1}\,(C_i+D_i)
\end{equation}
Here, for example, we give a few first $D_i$'s:
\begin{eqnarray}
D_1 & = & 0 \nonumber \\
D_2 & = & -A_1 x_1 +x_1 B_1  \\
D_3 & = & -A_1 x_2 -A_2 x_1 +x_2 B_1 +x_1 B_2
-\frac{1}{2}C_1 x_1^Tx_1 -\frac{1}{2}x_1x_1^TC_1
-x_1 C_1^T x_1 +\frac{1}{2} x_1x_1^TA_0x_1\,. \nonumber
\end{eqnarray}
This completes the block diagonalization of the symmetric matrix $S$.
To summarize, assuming all the $x_j$'s are known up to $j=i-1$,
$x_i$ can be easily obtained by solving
$(\tilde{S}_{12})_{i} ={\bf 0}_{n\times m}$ and then
$(x_1,x_2,...,x_i)$
fixes $y_{i+1}$ and $z_{i+1}$ for the mixing matrix
and the block-diagonalized matrices up to the $\epsilon^{i+1}$ order.

As a simple application of our method, we consider the situation
\footnote{See, also, the Appendix in Ref.~\cite{Miller:2003ay}.}
\begin{equation}
A=A_0\,, \ \ \ B=\epsilon^2 B_2\,, \ \ \ C=\epsilon C_1\,.
\end{equation}
In the first order of $\epsilon$, 
\begin{equation}
x_1=A_0^{-1} C_1 \,, \ \ \
\end{equation}
which leads to
\begin{equation}
y_2=-\frac{1}{2} x_1 x_1^T\,, \ \ \
z_2=-\frac{1}{2} x_1^T x_1\,, \ \ \
\end{equation}
for the mixing matrix and the block-diagonalized matrices are
given by
\begin{eqnarray}
\tilde{S}_{11} =
A_0+\frac{1}{2}\epsilon^2 (C_1x_1^T+x_1 C_1^T)\,;
\ \ \ \
\tilde{S}_{22}& =& \epsilon^2 (B_2 -C_1^T x_1)\,.
\end{eqnarray}
Note that, in this simple case, $x_2$ and, accordingly, $y_3$ and $z_3$ vanish.

%%%%%%%%%%%%%%%%%%%%%%%%%%%%%%%%%%%%%%%%%%%%%%%%%%%%%%%%%%%%%%%%%%%%%%%%%
\end{appendix}

%\newpage

%
%

\end{document}